\documentclass{article}[11pt]

\usepackage{mathtools,mathrsfs,amssymb,amsfonts,graphicx,setspace}

\usepackage{xcolor}

\usepackage{booktabs} 
\usepackage{array}    
\newcolumntype{L}{>{$}l<{$}} 
\newcolumntype{R}{>{$}r<{$}} 
\newcolumntype{C}{>{$}c<{$}} 

\usepackage{cite} 

\usepackage{slashed}

\usepackage{color,longtable}

\definecolor{blue-violet}{rgb}{0.54, 0.17, 0.89}
\definecolor{PineGreen}{cmyk}{0.92, 0, 0.59, 0.25}
\definecolor{OliveGreen}{cmyk}{0.64, 0, 0.95, 0.40}
\definecolor{RawSienna}{cmyk}{0, 0.72, 1, 0.45}
\definecolor{Gray}{cmyk}{0, 0, 0, 0.50}
\definecolor{MidnightBlue}{cmyk}{0.98, 0.13, 0, 0.43}
\definecolor{Orange}{cmyk}{0, 0.61, 0.87, 0}
\definecolor{LimeGreen}{cmyk}{0.50, 0, 1, 0}
\definecolor{Green}{cmyk}{1, 0, 1, 0}

\newcommand{\commas}{``}

 \newcommand{\diff}{\mathrm{d}}

\usepackage[makeroom]{cancel}
\usepackage[normalem]{ulem}

\usepackage{amsmath,amssymb,graphicx}
\usepackage{amssymb,graphicx}

\usepackage{amsfonts,amsmath}
\usepackage{latexsym}

\def\cN{{\cal N}}

\def\id{{\mathbb{I}}}

\def\cN{{\cal N}}

\def\dM{{\partial\mathcal{M}_4}}

\def\N8{{\cN\!=\!8}}

\numberwithin{equation}{section}
\allowdisplaybreaks

\begin{document}

\begin{center}
{\bf \Large  $D=4,\,\cN=2$ Supergravity: \\\vskip 5mm An Unconventional Application}\\[7mm]
  Laura Andrianopoli\\[4mm]
\noindent
{\small
{Politecnico di Torino, Corso Duca degli Abruzzi 24, 10129 Torino, Italy}\\
 {INFN, Sezione di Torino, Via Pietro Giuria 1, 10125 Torino, Italy}
\\[1mm]
 Email: {\tt laura.andrianopoli@polito.it}} \\
\vskip 1cm
{\bf{Abstract:}}
\end{center}
I will report on a top-down approach relating $\mathcal{N}=2$, $D=4$ pure supergravity with non-trivial boundary behaviour to a $(2+1)$-dimensional analog model which is able to describe the electronic properties of graphene-like materials. This is obtained, in a special asymptotic limit, by imposing an unconventional realization of supersymmetry in the $D=3$ boundary model.

\vfill

\begin{center}
\noindent
Invited contribution to the book:\\ {\it Half a Century of Supergravity}, eds. A. Ceresole and G. Dall'Agata \\(Cambridge University Press, to appear)
\end{center}

\thispagestyle{empty}
\newpage
\section{Introduction}
The present contribution aims to report on a series of papers regarding an \commas unconventional" use of Supergravity, where the radial component of the gravitino field at the asymptotic boundary serves, in an analog way, as the single-electron  wave-function  in a graphene-like sheet, the starting point being asymptotically AdS$_4$, pure $\mathcal{N}=2$ supergravity in $D=4$ space-time dimensions (I will refer to it in the following as AdS$_4$ supergravity).\par
This  application is \emph{unconventional} because it does not fall, in its present form, into a standard well-assessed holographic duality framework \cite{Maldacena:1997re,Gubser:1998bc,Witten:1998qj}, but also because it crucially makes use of an unconventional realization of supersymmetry on the three-dimensional boundary,  introduced in \cite{Alvarez:2011gd} and named by its authors \emph{matter Ansatz}.

To establish the holographic relation between AdS$_4$ supergravity  and graphene-like materials \footnote{By  \commas graphene-like materials" we mean (non-relativistic) materials in two spatial dimensions sharing with graphene the feature of having a honeycomb lattice, and then a bipartite single-electron wavefunction behaving, near the Dirac points,   as a Dirac fermion.  } \cite{Novoselov66}, requires a set of non-trivial steps:\par
The first step is to identify an appropriate asymptotic limit relating AdS$_4$ supergravity,  with asymptotically non-vanishing fields, with its $D=3$ asymptotic boundary theory. This will be the subject of Section \ref{aads4}, where I will review the main properties of AdS$_4$ supergravity, focussing in particular on the case where the asymptotic boundary behavior of the fields is non-trivial, following \cite{Andrianopoli:2014aqa}, and I will present, following \cite{Andrianopoli2018}, the special asymptotic limit leading to a  family of $D=3$ supergravities,  known as \commas Achucarro-Townsend" (A-T) supergravities \cite{Achucarro:1987vz,Achucarro:1989gm}, which in our model should be located on a locally AdS$_3$  patch of the asymptotic boundary. \par
The next step is the implementation of the matter Ansatz of \cite{Alvarez:2011gd} in the resulting asymptotic   $\mathcal{N}=2$ A-T model. As we will see, the resulting model features a massive propagating Dirac spinor, whose mass is sourced by the space-time torsion.
This will be presented in
 Section \ref{3d}, where I will first describe the main properties of A-T supergravities, and I will then report in some detail the prescriptions of unconventional supersymmetry \cite{Alvarez:2011gd}, together with the consequences, first discussed in \cite{Andrianopoli2018}, of imposing the corresponding matter Ansatz when the model is embedded in  the $D=4$ supergravity theory.
 \par
 The last step is the definition of the analog relation with graphene.
This requires some more effort because the symmetry of the model obtained from the above steps does not feature enough symmetry to get a realistic description as the single-electron wave function of graphene. A suitable analog formulation was obtained in \cite{Andrianopoli:2019sip}, by embedding the $\mathcal{N}=2$ model in an $\mathcal{N}=4$ A-T supergravity (and then in the corresponding $\mathcal{N}=4$ bulk supergravity), thus providing the quantum numbers required to accomodate a possible graphene wave-function. This will be presented in Section \ref{graf}, where I will be ready to describe, following \cite{Andrianopoli:2019sip}, the main features of the analog model and to make contact with the electronic properties of graphene-like materials.

Throughout the paper we use a mostly minus space-time signature, both in four and in three dimensions.

\section{Pure $\mathcal{N}=2$, asymptotically AdS$_4$ supergravity\\and its boundary behavior}
\label{aads4}
In this section, following \cite{Andrianopoli:2014aqa}, I will review the geometric construction of  pure $\mathcal{N}=2$  supergravity, in the presence  of  non-trivial boundary conditions on the fields involved.

The idea of supergravity   was first introduced  by S. Ferrara, D. Freedman and P. Van Nieuwenhuizen in 1976 \cite{Freedman:1976xh} and soon after, independently, by S. Deser and B. Zumino \cite{Deser:1976eh}. \footnote{The  number of important contributions to the construction of Supergravity is really huge. Due to the limited length of the contribution,  and to the worry of forgetting some fundamental contribution,  I decided not to cite here any of the foundational papers on Supergravity besides the first ones and those strictly related to the research presented here.}
It is an extension of General Relativity, endowed with an extra continuous symmetry structure associated with  Grassmann-odd generators, that is supersymmetry.
Soon after its formulation, it was realized that the peculiar symmetry structure of supersymmetry is very well accounted for in the so-called Superspace
approach  \cite{Neeman:1978njh,Sohnius:1985qm,Wess:1992cp}. This corresponds to extend the notion of space-time to that of Superspace, which is an enlarged space parametrized by the ordinary  space-time coordinates $x^\mu$, together with Grassmann-odd coordinates $\theta^\alpha$. In this framework, supersymmetry acts as a superdiffeomormophism, that is  as a diffeomorphism along the odd directions of superspace. 
In the present review I will adopt a superspace approach, and in particular the geometric \emph{rheonomic} approach, introduced in \cite{Neeman:1978njh} and fully developed by L. Castellani, R. D'Auria and P. Fr\'e in \cite{Castellani:1991et}, which is the supersymmetric extension of the Einstein-Cartan formulation of General Relativity \cite{Cartan:1923zea}.

In the $\mathcal{N}$-extended theories, the Fierz identities among the gravitini   are a source of nontrivial cocycles in superspace, thus endowing these models with a much richer cohomological structure with respect to the minimal case, responsible for the central extension of the supersymmetry algebra (with $\mathcal{N}>1$), where the central charges are, in fact, \emph{topological charges} \cite{Witten:1978mh}.
 This is associated with the fact that the field content of the gravitational multiplet in the extended cases includes gauge fields, the graviphotons, and this naturally endows the corresponding theories with the action of the flat symplectic bundle of electric/magnetic duality transformations.

Among the $\mathcal{N}$-extended theories in four-dimensional space-time,
$\mathcal{N}=2$ Supergravity is  the least constrained supersymmetric dynamical model, still featuring a nontrivial cohomological structure of superperspace.
 In this case, the field content is given, besides the metric and a couple of gravitini, by a single graviphoton field, $A_\mu$.

The dynamics of the supergravity theories is encoded in the  \emph{gauging}, corresponding to switch on minimal coupling between (a subset of) the gauge fields of the theory and the rest of  supergravity and matter fields, thus introducing mass terms and a scalar potential. The possible gauge groups are subgroups of the scalar-sigma-model isometry groups, and the different dynamical models follow by the different choices of them among the possible ones compatible with supersymmetry.
Even in the absence of matter, which in the $\mathcal{N}=2$ case implies the absence of scalar fields at all,  it is possible to switch-on a sort of gauging, through a so-called Fayet-Iliopoulos (FI) term. In this case, the scalar potential reduces to a constant,  the cosmological constant.  Supersymmetry preserving solutions require it to be  negative, and the coresponding supergravity theory is AdS$_4$ supergravity, describing fluctuations around a super anti-de Sitter (AdS) background.
This model does not include fields of spin lower than 1, so that the theory can be formulated fully geometrically in terms of 1-form fields and their supercurvatures.

AdS$_4$ geometry has an asymptotic boundary, located at radial infinity. In general, the presence of a space-time defect, or of a boundary, introduces a length scale in the theory, thus spoiling the invariance under space-time diffeomorphisms of gravitational theories. Diffeomorphism invariance can be recovered by including appropriate boundary terms in the Lagrangian, as first observed in  \cite{York:1972sj,Gibbons:1976ue}. This problem was then  afforded by many authors in different contexts, and within a frame-independent approach (à la Cartan), an interesting result was obtained
 in  \cite{Aros:1999id,Olea:2005gb,Miskovic:2009bm} for $D=4$  asymptotically AdS theories. For these theories, in \cite{Aros:1999id,Olea:2005gb,Miskovic:2009bm} it was shown that the boundary term to be added is the topological  Euler-Gauss Bonnet term, which regularizes the Action and leads to a background-independent definition of conserved charges.
 At the supergravity level, this problem was considered by several authors (see, e.g., \cite{vanNieuwenhuizen:2005kg,esposito,Amsel:2009rr}) and in a frame-independent way, for  asymptotically super-AdS$_4$ $\mathcal{N}=1$ and $\mathcal{N}=2$ supergravity, in \cite{Andrianopoli:2014aqa}. For asymptotically-flat supergravity, the construction in a frame-independent way is still possible \cite{Concha:2018ywv}, but it requires the addition of \commas topological" auxiliary fields. \footnote{For a review on the geometric approach to the boundary problem in supergravity, see \cite{Andrianopoli:2021rdk} which also contains an extensive list of references on the subject.}

The starting point here is AdS$_4$ supergravity. As I am going to review, in this case the  boundary terms needed to recover supersymmetry
invariance of the action are the $\mathcal{N}=2$ supersymmetric extension of the  Euler-Gauss-Bonnet term.

\subsection{The bulk theory}

The general expression, in the  bulk, of pure $\mathcal{N}=2$ AdS$_4$ supergravity can be found from the general matter-coupled $\mathcal{N}=2$ gauged supergravity of \cite{Andrianopoli:1996vr,Andrianopoli:1996cm}, in the absence of matter multiplets, still keeping a FI term as a constant flux. This introduces a length scale in the Lagrangian  associated with the AdS radius $\ell$, providing   a constant  mass-matrix to the gravitini and a negative cosmological constant $\Lambda= -\frac 3{\ell^2}$.
The choice of the FI term, pointing in a specific $SO(2) \subset SU(2)_R\times U(1)_R$ direction,    breaks the R-symmetry invariance of the theory to $SO(2)$.
More precisely, referring to the formulation in \cite{Andrianopoli:1996cm}, in this case the (complex) gravitino mass-matrix $S_{AB}$
reduces to a real (or purely imaginary) tensor, which can be chosen as 
 $S_{AB}= -\frac 1{2\ell} \delta_{AB}$.
The field content of the theory is given by the bosonic vielbein 1-form $V^a$, a couple of Majorana-spinor gravitini 1-forms $\Psi_{A}$ (with  $A=1,2$  in the fundamental representation of $\mathrm{SO}(2)$), the Lorentz spin connection $\omega^{ab}$, and the graviphoton $A $, which is an Abelian gauge connection.
Our conventions on $D=4$ fermions can be found in the Appendices of \cite{Andrianopoli:1996cm,Andrianopoli:2014aqa}. \footnote{Here,  the notation for the spin-connection is different from the one in  \cite{Andrianopoli:1996cm, Andrianopoli:2014aqa}. In particular, with respect to those papers: $\omega^{ab}\rightarrow -\omega^{ab}$. Furthermore, here we chose
 as graviphoton
 gauge field $A$,   the \commas dressed" connection $A\equiv Im(\mathcal{N}_{00})L^0 A^0= -L^0 A^0$. This is consistent with the relation  of special geometry $Im(\mathcal{N}_{00})|L^0|^2= -\frac 12$, which implies in this case $|L^0|^2=\frac 12$. }


The bulk Lagrangian 4-form in $\mathcal{N} = 2$ superspace is written in terms of the $\mathrm{SO}(1,3)\times \mathrm{SO}(2)$-covariant supercurvatures in superspace:
\begin{subequations}\label{curvN2}
\begin{align}    \mathcal{R}^{ab}& \equiv \diff \omega^{ab} + \omega^{ac} \wedge {\omega_c}^b \,,\hfill \\
    T^a & \equiv \mathcal{D} V^a -\frac{i }{2}\bar\Psi_A \Gamma^a \wedge \Psi_A\,,\hfill\\
    \rho_A & \equiv \mathcal{D} \Psi_A - \frac{1}{2\ell} A \epsilon_{AB} \wedge \Psi_B  \,, \\
    F & \equiv  \diff A - \bar \Psi_A \wedge \Psi_B\epsilon_{AB} =  \mathcal{F} - \bar \Psi_A \wedge \Psi_B\epsilon_{AB} \,.\hfill
\end{align}
\end{subequations}
and it reads
\cite{Andrianopoli:2014aqa,Andrianopoli:2020zbl}
\begin{align}
\mathcal{L}_{\text{bulk}} & =  \frac{1}{4} \mathcal{R}^{ab}\wedge V^c\wedge V^d\epsilon_{abcd}+\bar \Psi_A\Gamma_a\Gamma_5 \wedge \rho_{A}\wedge V^a+\nonumber\\
& +\frac 14 \epsilon_{abcd}\tilde F^{cd} V^a\wedge  V^b
\wedge F -\frac 1{48} \tilde F_{\ell m}\tilde F^{\ell m}  V^a\wedge V^b \wedge V^c \wedge V^d \epsilon_{abcd}+\nonumber\\
&+\frac i2\left( F+\frac 12 \bar \Psi_A \wedge \Psi_B\epsilon_{AB}\right) \wedge \bar\Psi _C\Gamma_5\wedge \Psi_D\epsilon_{CD} +
 \nonumber\\
&-\frac{i }{2\ell}\bar\Psi_A\Gamma_{ab}\Gamma_5\wedge \Psi_{A}\wedge V^a\wedge V^b-\frac{1}{8\ell^2}V^a\wedge V^b\wedge V^c\wedge V^d\epsilon_{abcd} \label{LbulkN2}
\end{align}
where we have written the only component of the gauge kinetic matrix $\mathcal{N}_{\Lambda\Sigma}$ of the graviphoton field strength, which in this case is a constant,   as $\mathcal{N}_{00}=\Theta -i $ (the minus sign is required for a proper gauge kinetic term).
Note that  the terms proportional to $\Theta$ in the Lagrangian sum up to a boundary term, $\Theta$ being in this case a constant theta-term, and therefore they are not included in the bulk Lagrangian \eqref{LbulkN2}.
In the following, I will   omit the \commas wedge" symbol in the product of differential forms.

Following the prescriptions of the geometric approach \cite{Castellani:1991et}, the Lagrangian \eqref{LbulkN2} is written at {first} order for the spin connection $\omega^{ab}$, as is usual in gravitational theories à la Cartan, but also for the graviphoton gauge field $A$. This is necessary in order to write a fully geometric and frame-independent Lagrangian in superspace, which implies  avoiding use of the metric field. In this way, we can define the Action of the theory  by the integral of the 4-form Lagrangian over any bosonic four-dimensional Minkowskian submanifold, $\mathcal{M}_4$, immersed in the $\mathcal{N}=2$ superspace $\mathcal{M}^{(4|8)}$:
 \begin{equation}
     \mathcal{A}= \int_{\mathcal{M}_4\subset \mathcal{M}^{4|8}}\mathcal{L}\,.\label{action}
 \end{equation}
 Then, the Euler-Lagrange equations, holding in the full superspace, are obtained from the Action principle $\delta \mathcal{A}=0$, and they are independent on the choice of the particular hypersurface of integration $\mathcal{M}_4$ since, the lagrangian \eqref{LbulkN2}   being geometric in superspace, any variation of the domain of integration can be reabsorbed in a diffeomorphism in superspace. \footnote{\label{refrheo}The general discussion of this issue is given in the book \cite{Castellani:1991et} and, for the specific theory under consideration, in \cite{Andrianopoli:2014aqa}. More recent reviews discussing this point in some detail are  \cite{DAuria:2020guc} and \cite{Andrianopoli:2024qwm}. }

 The relation of the first-order Lagrangian \eqref{LbulkN2} with the standard second-order formulation is obtained by its variation with respect to the auxiliary (super)fields $\tilde{F}_{ab}$ and $\omega^{ab}$.
 Variation with respect to the tensor 0-forms $\tilde{F}_{ab}$ enforces the condition $\tilde{F}_{ab}=F_{ab}$,  identifying it with the supercovariant field-strength $F_{ab}$, that is the component along the purely bosonic vielbein of the field-strength 2-form $F$. On the other hand, the field equation of the spin connection gives the on-shell constraint of
 vanishing supertorsion
\begin{equation}
     {T}^{a} \equiv \mathcal{D} V^a -\frac{i }{2}\bar\Psi_A \Gamma^a \Psi_A =0\,.
\end{equation}

\subsection{Supersymmetry invariance}
In the geometric  superspace approach, the supersymmetry transformations of the (super)fields are nothing but diffeomorphisms along the odd-directions of superspace. As such,  they can be  determined through Lie derivatives.
Given a generic superfield, $\Phi$, we have therefore:
\begin{equation}
    \delta_\epsilon \Phi=\ell_\epsilon \Phi =   \diff \iota_\epsilon (  \Phi )
    +\iota_\epsilon\,  (\diff\Phi)\,,\label{liephi0}
\end{equation}
which can be rewritten, in terms of covariant derivatives $\nabla$, as:
\begin{equation}
    \delta_\epsilon \Phi=\ell_\epsilon \Phi =   \nabla \iota_\epsilon (  \Phi )
    +\iota_\epsilon\,  (\nabla\Phi)\,.\label{liephi}
\end{equation}
Here, $\iota_\epsilon$ denotes the contraction operator along an odd direction of superspace, $\epsilon \equiv \bar \epsilon_A D^A $ being the parameter corresponding to an infinitesimal supersymmetry transformation, and $\nabla\Phi$ is the fieldstrength of the given superfield, that is one of the supercurvatures \eqref{curvN2}. The odd directions are given in terms of the spinorial tangent vectors ${D^A}$, which are dual to the gravitino 1-forms:
$\bar\Psi_A(D^B)=\delta^B_A$, so that:
\begin{align}
 \iota_\epsilon ( \Psi_A)= \epsilon_A \,,\quad \iota_\epsilon ( V^a)= 0\,,\quad \iota_\epsilon ( \omega^{ab})= 0\,,\quad \iota_\epsilon ( A)= 0\,.
\end{align}
There is a subtlety here, in evaluating the supersymmetry transformations of the fields:
 in calculating the contraction of the   supercurvatures in \eqref{liephi} one has to use, instead of the formal definitions \eqref{curvN2}, the rheonomic parametrizations 
of the supercurvatures  as   2-forms in superpace:
\begin{subequations}
\begin{align}
\mathcal{R}^{ab}=& \tilde{\mathcal{R}}^{ab}{}_{cd}V^c V^d + \bar \Theta^{ab|c } _A \Psi_A V_c  +\frac 1{2\ell} \bar\Psi_A \Gamma^{ab}\Psi_B\delta_{AB}+\hfill\nonumber\\
&-\frac 12 \epsilon_{AB}\left(\tilde{F}^{ab}\bar\Psi_A \wedge \Psi_B+\frac i2 \epsilon^{abcd}\tilde{F}_{cd}\bar\Psi_A \Gamma_5\wedge \Psi_B
\right)\,, \\
  {T}^{a}=& 0\,,\hfill\\
  \rho_A  =&\tilde\rho_{ab} V^a V^b
  -\frac i 2  \epsilon_{AB} \left(\tilde F_{ab} -\frac i2 \epsilon_{abcd}\tilde{F}^{cd}\Gamma_5\right)\Gamma^a \Psi_BV^b +\nonumber\\
 & + \frac i { 2\ell} \Gamma_a \Psi_A V^a\,, \hfill\\
    F   =&  \tilde F_{ab}V^a V^b \,,
\end{align}\label{curvN2par}
  \end{subequations}
where $\tilde{\mathcal{R}}^{ab}{}_{cd},\,\tilde\rho_{ab} ,\,\tilde F_{ab}$ are the  supercovariant field-strengths, and $\bar \Theta^{ab|c} _{A}  =2i \Gamma^{[a}\tilde\rho_A^{b]c}-i \Gamma^{c}\tilde\rho_A^{ab}$. The para\-me\-trizations \eqref{curvN2par} hold on-shell.

This  crucial requirement  implements in the geometric approach  the condition that supersymmetry is realized on the  superfields  as an on-shell symmetry. It incorporates the principle of rheonomy and guarantees the equivalence of diffeomorphisms in the odd directions of superspace with supersymmetry transformations of the fields on spacetime.
  An extended  discussion of this point, and of the rheonomy principle, is given in the book \cite{Castellani:1991et}.
 Some recent reviews on the subject  are cited in footnote \ref{refrheo}.
\par
The supersymmetry transformation of the Action \eqref{action} in superspace is:
\begin{align}\label{lielag}
    \delta_\epsilon \mathcal{A}= \int_{\mathcal{M}_4\subset \mathcal{M}^{4|8} }\ell_\epsilon  \mathcal{L} = \int_{\mathcal{M}_4} \Bigl(d\iota_\epsilon ( \mathcal{L})+\iota_\epsilon (d \mathcal{L})\Bigr)\,.
\end{align}
The first contribution in \eqref{lielag} is a boundary term, not constraining the bulk.
As for the second contribution, it
would be a trivially vanishing term if $\mathcal{L}$ would be a 4-form in four-dimensional space-time (in that case  $\mathcal{L}$ would be a top form, and then  $d\mathcal{L}=0$, being a 5-form). In superspace instead, where the 4-form $\mathcal{L}$  is not a top-form,  the condition \begin{equation}
    \iota_\epsilon(d\mathcal{L})=0\label{iotadl}
\end{equation}is non-trivial. It is the necessary condition to be satisfied for supersymmetry invariance of the bulk Lagrangian.

In the particular case of interest here,  the supersymmetric Lagrangian \eqref{LbulkN2} satisfies \eqref{iotadl}, up to boundary terms. 


 \subsection{Boundary contributions}
Let us consider here the case that the boundary behavior of the superfields in the bulk-supersymmetric Lagrangian  \eqref{LbulkN2} be non trivial.
In this case, the condition for supersymmetry invariance of the Action, eq. \eqref{lielag}, reduces to:
\begin{align}
    \delta_\epsilon \mathcal{A}=   \int_{\mathcal{M}_4}  d\iota_\epsilon ( \mathcal{L})= \int_{\partial\mathcal{M}_4} \iota_\epsilon ( \mathcal{L})=0\,.
\end{align}
A consistent definition of the supersymmetric action in this case requires to modify the lagrangian by including boundary contributions, namely
\begin{equation}
\mathcal{L}_{\text{bulk}}\to\mathcal{L}_{\text{full}} \equiv \mathcal{L}_{\text{bulk}} + \mathcal{L}_{\text{bdy}} \,,\label{Lfull}
\end{equation} with $\mathcal{L}_{\text{bdy}}$ such that
\begin{align}
    \iota_\epsilon (\mathcal{L}_{\text{full}})\bigr\vert_{\partial\mathcal{M}_4}=0\,.\label{deltaLfull}
\end{align}
For the case of $\mathcal{N}=2$ pure supergravity, with bulk supersymmetric lagrangian   given by \eqref{LbulkN2}, the required boundary Lagrangian    has the form  \cite{Andrianopoli:2014aqa}
\begin{align}\label{LbdyN2EGB}
\mathcal{L}_{\text{bdy}} = -\frac{\ell^2}{8}\Bigl( &\mathcal R^{ab} \mathcal R^{cd}\epsilon_{abcd}+\frac{8{i }}{\ell} \bar{\rho}_A\Gamma_5 \rho_{A}+\nonumber \\
&-\frac{2 i }{\ell}\mathcal R^{ab}\bar\Psi_A\Gamma_{ab}\Gamma_5\Psi_{A}+\frac{4 i }{\ell^2}dA \bar\Psi_A\Gamma_5\Psi_B\epsilon_{AB}\Bigr) \,.
\end{align}
which is the $\mathcal{N}=2$ supersymmetric extension of the purely bosonic Euler-Gauss-Bonnet term \cite{Aros:1999id,Olea:2005gb,Miskovic:2009bm}.
We note in particular that supersymmetry invariance of $\mathcal{L}_{\text{full}}$ does not allow for a theta-term in the gauge sector.

Taking into account the boundary contributions, the supersymmetric 4-form Lagrangian $\mathcal{L}_{\text{full}}$ is then the sum of the bulk contribution \eqref{LbulkN2} and the boundary one, \eqref{LbdyN2EGB}.
 The space-time Lagrangian can be straightforwardly  derived from it as a projection. See \cite{Castellani:1991et} for a general discussion, or Appendix A of Ref. \cite{Andrianopoli:2014aqa} for the details in the present case.

The superspace Euler-Lagrange equations of
\eqref{Lfull} are now  extended to hold in the full superspace (including its boundary), and they imply the following boundary constraints to be satisfied by the supercurvatures:
\begin{subequations}\label{osp0curv}
\begin{align}
\mathcal{R}^{ab}\bigr\vert_{\partial \mathcal{M}_4} = &\left[ \frac{1}{\ell^2} V^a V^b + \frac{1}{2\ell} \delta^{AB} \bar\Psi_A \Gamma^{ab}\Psi_B \right]_{\partial \mathcal{M}_4}\,,\hfill \\
\mathcal{D}V^a\bigr\vert_{\partial \mathcal{M}_4} = & \left[ \frac{i }{2}\bar\Psi_A \Gamma^a \Psi_A\right]_{\partial \mathcal{M}_4} \,,\hfill \\
\rho_A\bigr\vert_{\partial \mathcal{M}_4} = &   \left[\frac{i }{2\ell}\delta_{AB}\Gamma_a \Psi_B V^a\right]_{\partial \mathcal{M}_4} \,,\hfill \\
 F\bigr\vert_{\partial \mathcal{M}_4} = &0 \,.\hfill
\end{align}
\end{subequations}
 The above constraints are nothing but the conditions for asymptotically global $\mathrm{OSp}(2|4)$ invariance of the Action.
 This can be rephrased  as the condition that the superfields do satisfy asymptotically the Maurer-Cartan equations defining the $OSp(2|4)$ superalgebra in its Poincaré dual form:
\begin{subequations}\label{lagsuperN2}
\begin{align}
\mathrm{R}^{ab}\equiv&\diff \omega^{ab} + \omega^{ac} {\omega_c}^b - \frac{1}{\ell^2} V^a V^b - \frac{1}{2\ell} \delta^{AB} \bar\Psi_A \Gamma^{ab}\Psi_B \to 0 \,, \hfill\\
\mathrm{T}^{a}\equiv&\diff V^a +\omega^{ab} V_b - \frac i2 \bar\Psi_A \Gamma^a \Psi_A = 0 \,, \hfill \\
\hat{\rho}_A\equiv&\diff \Psi_A +\frac{1}{4} \omega^{ab}\Gamma_{ab} \Psi_A- \frac{1}{2\ell} A \epsilon_{AB} \Psi_B  -    \frac{i }{2\ell}\delta_{AB}\Gamma_a \Psi_B V^a \to 0\,,\hfill \\
F\equiv&\diff A - \bar \Psi_A \Psi_B\epsilon_{AB} \to 0 \,.\hfill
\end{align}
\end{subequations}
Eq.s \eqref{lagsuperN2} are the conditions for a locally asymptotically super-AdS$_4$ geometry or,  when expressed in terms of
the boundary superfields, the conditions for global $\mathcal{N}=2$ superconformal invariance of the boundary theory.

\subsection{A special asymptotic limit} \label{limit}

Asymptotically AdS geometries have been deeply investigated in the last 30 years, being at the heart of the AdS/CFT duality \cite{Maldacena:1997re,Gubser:1998bc,Witten:1998qj}. This celebrated holographic  duality  relates the classical behavior of locally asymptotically AdS (super)gravity to that of a quantum field theory  in one dimension less, actually a conformal field theory in the original formulation.
In the duality,
the radial variable   of  supergravity near the boundary is related to the energy scale in the dual theory, which can be thought of as living at the asymptotic  boundary. To handle the duality, the powerful tool of holographic renormalization was developed (see, for example, \cite{Skenderis:2002wp,DeWolfe:2018dkl} and references therein). This makes use of an asymptotic expansion in the radial coordinates, known as  Fefferman-Graham (FG) parametrization.
A detailed analysis of the holographic renormalization scheme to the subject of the present section, that is $\mathcal{N}=2$ AdS$_4$ supergravity,  with all the contributions, including the fermionic ones,  can be found in  \cite{Andrianopoli:2020zbl}.

I will not elaborate further on the details of the duality, here,  focussing instead on the gravity side of the correspondence only. I  would like to report on a special asymptotic limit of asymptotically AdS$_4$ supergravity, whose boundary behavior  will be the object of next sections.
As we have seen, the asymptotical behaviour \eqref{lagsuperN2} of the supercurvatures is dictated by the supersymmetry invariance of the Action.
Being expressed in terms of the superfields of the $D=4$ theory, this leaves room for investigating   a collection of possible specific asymptotics once, close to the boundary, the superfields are expressed as   explicit power series in the radial coordinate of superfields living at the boundary.
This explicit $D=3$ description depends on the specific properties of the theory at the boundary $\partial\mathcal{M}_4$ which we wish to relate to, inside the conformal class of possible geometries.

The (2+1)-dimensional  asymptotic geometry I am interested in here, at the boundary of $\mathcal{N}=2$ AdS$_4$ supergravity, is a locally AdS$_3$ space-time, since this is
the proper geometry
to make contact with a (2+1)-dimensional model introduced in \cite{Alvarez:2011gd} and featuring an unconventional realization of supersymmetry allowing to describe, in an analog way, some electronic properties of graphene-like materials. 
To obtain a  locally AdS$_3$ space (which itself has a boundary), as boundary geometry, we should impose 
a specific asymptotic behavior, on a patch of  asymptotically  AdS$_4$ space-time endowed with a locally AdS$_3$ boundary.  This special limit is an instance of the so-called ``ultraspinning limit'' of an AdS$_4$-Kerr black-hole, defined in  \cite{Caldarelli:2008pz, Caldarelli:2012cm,Gnecchi:2013mja}, see also the solution found in \cite {Klemm:2014rda}. More details, in relation with the present problem, can be found in \cite{Andrianopoli2018}.

To perform the calculation, following \cite{Andrianopoli2018} we first choose  a FG parametrization of the $D=4$ geometry breaking the manifest asymptotic space-time symmetry  $\mathrm{Sp} (4) \sim \mathrm{SO}(2,3)\to {\rm SO}(1,1)\times {\rm SO}(1,2)$,  ${\rm SO}(1,2)$ being interpreted as the Lorentz group on $\dM$ and ${\rm SO}(1,1)$ as dilation symmetry.
Near the boundary, located at $r\to \infty$, the four space-time coordinates $x^{\hat\mu} =(x^\mu,r)$, with $\hat\mu=(\mu,\hat{3})$, $\mu=0,1,2$, split into the three coordinates $x^\mu$ on $\dM$ and the radial coordinate $x^{\hat{3}}=r$, while the flat 4D Lorentz indices split as $a=(i,3)$, with $i=0,1,2$.
Then, we rewrite the supergravity fields and  their supercurvatures \eqref{osp0curv} in a manifest ${\rm SO}(1,1)\times {\rm SO}(1,2)$-covariant way.
 It is useful to introduce the extrinsic curvature 1-form  $\omega^{3i}\equiv h^i$, and the 1-form $\mathrm{SO}(1,2)$-vectors $E_\pm^i$, with opposite ${\rm SO}(1,1)$-grading, as follows:
\begin{align}
E_\pm^i\equiv \pm\,\frac{1}{2}\left(V^i\mp \ell\,h^i\right)\,.\label{defef}
\end{align}
Similarly, we choose a Gamma-matrix representation where $\Gamma^3$ is block-diagonal, and decompose the four-component  gravitini $\Psi_A$ by projecting them into  $\Gamma^3$-chiral projections, $\Psi_{A\pm}\equiv \frac 12\left(\id \mp\,i \Gamma^3\right)\Psi$, with opposite  ${\rm SO}(1,1)$-grading:
\begin{equation}
\Psi_{A}=\Psi_{+\,A}+\Psi_{-\,A}\,\,,\,\,\,\,\,\Gamma^3\Psi_{\pm\,A}=\pm i \,\Psi_{\pm\,A}\,.
\end{equation}
In terms of these new quantities,  the boundary values \eqref{osp0curv} of the supercurvatures read:
\begin{subequations}
\label{curvs3}
\begin{align}\mathcal{R}^{ij}&\equiv d\omega^{ij}+\omega^i{}_k\wedge \omega^{kj}=-\frac{4}{\ell^2}\,E_+^{[i}\wedge E_-^{j]}+\frac{1}{\ell}\,\bar{\Psi}_{+\,A}\Gamma^{ij}\Psi_{-\,A}\,,\\
dV^3&=-\omega^3{}_i\wedge V^i+\bar{\Psi}_{+\,A}\Psi_{-\,A}\,,\\
{\mathcal{D}}E_\pm^i&\equiv  dE_\pm^i+\omega^{i}{}_j\wedge E_\pm^j=\pm \frac{i }{2}\,\bar{\Psi}_{\pm A}\Gamma^i\Psi_{\pm A}\pm\frac{1}{\ell}\,E_\pm^i\wedge V^3\,,\\
{\mathcal{D}}{\Psi}_{\pm A}&\equiv d{\Psi}_{\pm A}+\frac{1}{4}\omega_{ij}\,\Gamma^{ij}\wedge{\Psi}_{\pm A}\\
&=
\mp \frac{i }{\ell}\,E_{\pm i}\,\wedge\Gamma^i{\Psi}_{\mp A}   -\,\epsilon_{AB}\,A^{(4)}\wedge {\Psi}_{\pm |B}\pm \frac{1}{2\ell}\,{\Psi}_{\pm A}\wedge V^3\,,\\
dA^{(4)}&=\epsilon_{AB}\bar{\Psi}_{A}\,\wedge\Psi_{B}=2\,\epsilon_{AB}\bar{\Psi}_{+A}\,\wedge
\Psi_{-B}\,.\end{align}
 \end{subequations}
The prescription requires that the component of the vielbein orthogonal to the boundary $\dM$ should vanish on  it: $V^3|_{\dM}=\lim_{r\to \infty} V^3 =0$. This implies, for consistency, the condition $dV^3|_{\dM}= 0$, that is\footnote{One can easily show, by inspecting the Bianchi identities, that the extrinsic curvature 1-form $\omega^{3i}$ cannot have a component along an odd directions of superspace.}:
\begin{equation}
\label{dv3}
\omega^3{}_i |_{\dM}= \left(h_{i|j}\,V^j\right)_{\dM}\,\mbox{  with } h_{i|j}=h_{j|i}\,; \quad \bar{\Psi}_{+\,A}\Psi_{-\,A}=0\,.
\end{equation}
The second condition in \eqref{dv3}, in particular, implies that only up to a half of the $D=4$ supersymmetries will be preserved as supersymmetries of the boundary theory.

Fosussing on the space-time dependence of the superfields, to obtain as boundary geometry  a locally AdS$_3$ geometry, such that to make contact with the model in \cite{Alvarez:2011gd},  amounts to choose the following boundary behavior for the $D=4$ fields \cite{Andrianopoli2018,Andrianopoli:2019sip,Andrianopoli:2020zbl}, which is compatible with \eqref{dv3} \footnote{This special asymptotic behavior was previosly considered, for the $\mathcal{N}=1$ case, in \cite{Amsel:2009rr}.}:
\begin{align}\label{beha}
E_+^i(x,r)&=\frac{r}{2\ell}\left[\,E^i(x)+\mathcal{O}(\frac{\ell^2}{r^2})\right]\,;\,\,\,E_-^i(x,r)=-\,\frac{\ell}{2r}\,E^i(x)+\mathcal{O}(\frac{\ell^2}{r^2})\,;\nonumber\\
 \omega^{ij}(x,r)&=\omega^{ij}(x)+\mathcal{O}(\frac{\ell}{r})\,;\quad A^{(4)}(x,r)=2\ell\,\varepsilon \, A_\mu(x)\,dx^\mu+\mathcal{O}(\frac{\ell}{r})\,;\nonumber\\
\Psi_{+\,A\, \mu}(x,r)&=\sqrt{\frac{r}{2\ell}}\left[\begin{pmatrix}
\psi_{A\,\mu}\cr {\bf 0}
\end{pmatrix}+\mathcal{O}(\frac{\ell}{r})\right]\,;\,\,\Psi_{-\,A\,\mu}(x,r)=\sqrt{\frac{\ell}{2r}}\left[\begin{pmatrix}{\bf 0}\cr\varepsilon\,\psi_{A\,\mu}\end{pmatrix}+\mathcal{O}(\frac{\ell}{r})\right]\,;\nonumber\\
V^3(r)&=\frac{\ell}{r}\left[dr+\mathcal{O}(\frac{\ell^2}{r^2})\right]\,.
\end{align}
Here, the 1-form  $E^i\equiv E^i_\mu (x)dx^\mu$ and $\omega^{ij}(x)\equiv \omega^{ij}_\mu dx^\mu$  are the dreibein 1-form and spin-connection on $\partial \mathcal{M}_4$,
 respectively,
and $\psi_{A\,\mu}(x)=(\psi_{A\,\mu\,\alpha})$, $\mu=0,1,2$, $\alpha=1,2$, the gravitini in $D=3$, associated with 2-component Majorana spinor 1-forms $\psi_A\equiv\psi_{A\,\mu} dx^\mu$. The possible choice of sign, $\varepsilon =\pm 1$, is introduced in the asymptotics of $A^{(4)}$ and $\Psi_{-A\mu}$ for later convenience.  As for the radial components, $\Psi_{\pm A \,3}$, of the $D=4$ gravitini, their boundary behavior
will be reported later, in Section \ref{backto4}, and determined in such a way to comply with the model of \cite{Alvarez:2011gd}.


 If we now perform the $\frac\ell r\to 0$ limit in \eqref{beha}, the leading contributions  to eq.s \eqref{curvs3}, which define the relevant supercurvatures on $\dM$, are:
\begin{subequations}
\label{asymptoticeq}\begin{align}
\mathcal{R}^{ij}&=\frac{1}{\ell^2}\,E^{i}  E^{j}+\frac{\varepsilon}{2\ell}\,\bar{\psi}_A\gamma^{ij}\psi_A\,,  \\
{\mathcal{D}}E^i&=\frac{i }{2}\,\bar{\psi}_{A}\gamma^i\psi_{A}\,,  \\
{\mathcal{D}}\psi_A&=-\varepsilon\,
\frac{i }{2\ell}\,E_i\,\gamma^i  \psi_A -\,\epsilon_{AB}\,A \,  {\psi}_{B}\,, \\
dA&=-\frac\varepsilon{\ell}\epsilon_{AB}\bar{\psi}_{A}\psi_{B}\,,
\end{align}
\end{subequations}
where we denote from  now on with an abuse of notation, with the same symbols $\mathcal{R}^{ij}$ and $\mathcal D$ that denoted the Riemann tensor and Lorentz covariant derivative in $D=4$, the
corresponding quantities in $D=3$, expressed in terms of the spin connection $\omega^{ij}(x)$.

All together, eq.s \eqref{asymptoticeq} are the Maurer-Cartan equations of   ${\mathcal N}=2$ super AdS$_3$ algebra. They  can be also obtained  as Euler-Lagrange equations from the following $D=3$  lagrangian  \cite{Achucarro:1987vz}:
\begin{align}\label{lag3D}
 \mathcal{L}^{(3)}=& \left(\mathcal R^{ij} -\frac{1}{3\,\ell^2}  E^i E^j  -\frac{\varepsilon}{2\,\ell} \, \bar\psi_A\gamma^{ij}\psi_A\right) E^k \epsilon_{ijk} - \frac{\varepsilon}{2\,\ell} AdA +\nonumber\\
 &+2 \bar\psi_A\left(\mathcal{D} \psi_A - \frac{\varepsilon}{2\ell}\,\epsilon_{AB}\,A \,  {\psi}_{B}\right)\,.
\end{align}
Eq. (\ref{lag3D})  collects two inequivalent lagrangians in $D=3$ superspace, depending on the chosen sign of $\varepsilon=\pm 1$. They are instances of a    family of inequivalent ${\mathcal N}$-extended supergravity lagrangians in $D=3$,
introduced by Achucarro and Townsend in \cite{Achucarro:1987vz}, and  will be the subject of next section.

\section{Achucarro-Townsend Supergravity and \\Unconventional Supersymmetry} \label{3d}
  The possible unitary supersymmetry representations of locally AdS$_3$ supergravity
 in the  $\mathcal{N}$-extended case were classified in \cite{Gunaydin:1986fe}. They are associated with all the possible superalgebras whose bosonic subalgebra  includes $SO(2,2)$.
 However, supergravity theories in $D=3$ are non dynamical, since both the gravitational field and the gravitini carry on-shell zero local degrees of freedom. Still, it is possible to formulate non-trivial supergravity models in this non-dynamical case, which turn out to be espressible as Chern-Simons (CS, in the following) models.
 \subsection{Achucarro-Townsend supergravity}\label{ATsugra}
 In \cite{Achucarro:1987vz},
 the family of the ${\mathcal N}$-extended gauged supergravity lagrangians in $D=3$ based on the superalgebras $\mathfrak{osp}(p|2)_+ \times \mathfrak{osp}(q|2)_- $, with $p+q={\mathcal N}$,   was constructed by A. Achucarro and P.K. Townsend (these models will be referred to in the following as A-T supergravities).
 Each element of the family defines a  supersymmetric extension  of an AdS$_3$ background, and is  equivalent, up to boundary terms, to the CS Lagrangian of the same supergroup.

 The two $\mathcal{N}=2$, $D=3$  Lagrangian models \eqref{lag3D} labeled by  the parameter $\varepsilon=\pm 1$, introduced at the end of the previous section and whose Euler-Lagrange equations provide  the equations \eqref{asymptoticeq} defining the special limit connecting  $\mathcal{N}=2$ AdS$_4$ supergravity with its locally AdS$_3$ supersymmetric boundary, belong to the family of A-T supergravities, with superalgebras $\mathfrak{osp}(p|2)_{(\varepsilon)} \times \mathfrak{so}(1,2)_{(-\varepsilon)} $, where ($p =0, q=2$)  for the case
$\varepsilon=-1$, ($p =2,q=0)$
 for the case $\varepsilon=1$.
To make the invariance under the   superalgebra $\mathfrak{osp}(2|2)_{(\varepsilon)} \times {\mathfrak{so}}(1,2)_{(-\varepsilon)}$ fully manifest, it is convenient to switch to the CS reformulation of the model. This is obtained by introducing the torsionful connections
\cite{Achucarro:1989gm}:
\begin{equation}\label{omegavareps}
\omega_{(\pm\varepsilon)}^{ij} =  \omega^{ij} \pm \frac{\varepsilon}{\ell} \, E_k \epsilon^{ijk}\equiv\epsilon^{ijk}\,\omega_{(\pm\varepsilon)k}\,,
\end{equation}
 in terms of which the constraints (\ref{asymptoticeq}) read:
\begin{subequations}
\begin{align}
&\mathcal{R}^{i}_{(\varepsilon)}=i \, \frac{\varepsilon}{\ell}\,\bar{\psi}_A\gamma^{i}\psi_A\,, 
\quad
\mathcal{R}^{i}_{(-\varepsilon)}=0 \,,
\quad
dA = -\frac{\varepsilon}{\ell}\epsilon_{AB}\bar{\psi}_{A}\psi_{B}\, ,
\\
&{\mathcal{D}_{(\varepsilon )}}\psi_A=-\,\epsilon_{AB}\,A   {\psi}_{B}\,,
\label{asymptoticpsi}
\end{align}
\label{asymptotic}
\end{subequations}
where:
$ \mathcal{R}_{(\pm) i} \equiv \frac 12 \epsilon_{ijk}\mathcal{R}_{(\pm)} ^{jk}=d\omega_{(\pm) i} -\frac 12 \omega^j_{(\pm)}\omega^k_{(\pm)}\varepsilon_{ijk}$,
and $\mathcal{D}_{(\varepsilon)}$ denotes Lorentz-covariant derivative with the torsionful connection $\omega^{ij}_{(+\varepsilon)}$.
Eq.s \eqref{asymptotic} are indeed the Maurer-Cartan equations of the superalgebra $\mathfrak{osp}(2|2)_{(\varepsilon)} \times {\mathfrak{so}}(1,2)_{(-\varepsilon)}$ in its dual form.
\\
When written in terms of the supercurvatures  \eqref{asymptotic}, the supergravity lagrangian \eqref{lag3D} takes the form:
\begin{equation}
\label{SUGRA3D}
\mathcal{L}^{(3)} =\varepsilon\left( \mathcal{L}_{(\varepsilon)} - \mathcal{L}_{(- \varepsilon)}\right)+ d\mathcal{B}\,\equiv \mathcal{L}_+^{(3)}-\mathcal{L}_-^{(3)} + d\mathcal{B}
\end{equation}
 where:
 \begin{align}
\mathcal{L}_{(\varepsilon)}&=\,
\frac \ell 2\left( \omega_{(\varepsilon)}^i d\omega_{(\varepsilon)|i} -\frac 13  \omega_{(\varepsilon)}^i\omega_{(\varepsilon)}^j\omega_{(\varepsilon)}^k\varepsilon_{ijk} \right)
+2\varepsilon \,\bar \psi_A \nabla^{{(\varepsilon)}} \psi_A -\frac{\varepsilon}{2\ell} A\,dA \,,\label{supercs}\\
 \nabla^{(\varepsilon)} \psi_A&\equiv\,\left(d + \frac 14 \omega_{(\varepsilon)}^{ij} \gamma_{ij}\right)\,\psi_A +A \,\psi_B \,\epsilon_{AB} \,,\\
 \mathcal{L}_{(-\varepsilon)}&=\,
\frac \ell 2\left( \omega_{(-\varepsilon)}^i d \omega_{(-\varepsilon)|i} -\frac 13  \omega_{(-\varepsilon)}^i\omega_{(-\varepsilon)}^j\omega_{(-\varepsilon)}^k\varepsilon_{ijk} \right) \,,\\
\mathcal{B}&=-\frac \ell 2 \omega^i_{(+)}\omega_{(-) i}\,.
 \end{align}
 The 3-form
 $\mathcal{L}_{(+)}-\mathcal{L}_{(-)}$ is the CS lagrangian of the superalgebra $\mathfrak{osp}(2|2)_{(\varepsilon)}\times \mathfrak{so}(1,2)_{(-\varepsilon)}$, and it differs from the A-T lagrangian \eqref{lag3D} by the boundary term $d\mathcal{B}$.
The lagrangian (\ref{SUGRA3D}) is  invariant under the following supersymmetry transformations in superspace (written in terms of the connections \eqref{omegavareps}): \footnote{ Note that, since the theory is a CS model, here supersymmetry  is realized as a gauge symmetry.}
\begin{subequations}\label{susytrman}
\begin{align}
& \delta \omega^i_{(-\varepsilon)}=0\,, \quad \delta \omega^i_{(\varepsilon)}=\varepsilon\,\frac{2i }{\ell}\,\bar{\epsilon}_A\gamma^{i}\psi_A\,,\quad
\delta A = 2\epsilon_{AB}\bar{\epsilon}_{A}\psi_{B}\,, \\
&\quad\delta\psi_A={\mathcal{D}_{(\varepsilon)}}\epsilon_A\,-\frac{\varepsilon}{2\ell}\,\epsilon_{AB}\,A \,  {\epsilon}_{B}\equiv \nabla^{(\varepsilon)}\epsilon_A\,.
\end{align}
\end{subequations}
generated by the supersymmetry parameter $\epsilon_A$ associated with the tangent vector $\epsilon=\bar\epsilon_A D^A$ in the odd directions of $D=3$ superspace.

\subsection{The prescriptions of Unconventional Supersymmetry}\label{ususy}

 In  ref.~\cite{Alvarez:2011gd}, the $ \mathrm{OSp}(2|2)$-CS Lagrangian \eqref{supercs}, with an unconventional assumption on the Grassmann-odd part of the connection, was introduced.
A peculiarity of that model, which makes it particularly interesting, is that it provides an effective description of a   Dirac spinor satisfying a Dirac equation with mass related to the background torsion. As such, as we will see, it is particularly suited for describing, in an analog way,  electronic properties of graphene-like materials  
\cite{Iorio:2011yz,Guevara:2016rbl,Andrianopoli:2019sip}.

This Section aims to review, following \cite{Andrianopoli2018}, the main features of the  model in \cite{Alvarez:2011gd} and then to extend its prescriptions to the  supergravity model of Section \ref{ATsugra}, which was derived as a special asymptotic limit from $D=4$, $\mathcal{N}=2$ supergravity.
 However, to concretely apply the prescriptions of unconventional supersymmetry to an analog description as graghene-like systems requires a larger symmetry structure than the one of the models in \cite{Alvarez:2011gd,Andrianopoli2018}. This can be obtained by increasing the amount of supersymmetry. For this reason,  to make contact with graphene we will wait until the next section, where I will review, following \cite{Andrianopoli:2019sip}, the embedding of the above construction in a more general $\mathcal{N}$-extended pure supergravity.

The starting point of \cite{Alvarez:2011gd} is the 1-form connection $\mathbb{A}$ in the adjoint representation  of the supergroup  ${\rm OSp}(2|2)$:
\begin{equation} \label{A}
\mathbb{A}= \frac{1}{2}\,\omega^{ij}\, \mathbb{J}_{ij} + A\cdot \mathbb{T} + \overline{\psi}_A\,\mathbb{Q}^A + \overline{\mathbb{Q}}_A\,\psi^A\,,
\end{equation}
where $\omega^{ij}$, $A$ and $\psi^A$ are one-forms, while ${\rm \mathbb{J}}$, $\mathbb{T}$ and $\mathbb{Q}$  (with $\overline{\mathbb{Q}}= {\mathbb{Q}}^T\cdot C$) are the $\mathfrak{ osp}(2|2)$-generators of Lorentz, $U(1)$ and supersymmetry transformations, respectively.
This is the building block of the 3d CS Action
\begin{align}
\mathcal{A}= \int_{\mathcal{M}_3}\mathcal{L}(\mathbb{A})=\int_{\mathcal{M}_3}\left\langle\mathbb{A}\wedge d\mathbb{A}+ \frac 23 \mathbb{A}\wedge \mathbb{A}\wedge\mathbb{A}\right\rangle\label{acs}
\end{align}
where $\langle \cdots \rangle$ stands for the supertrace in $\mathfrak{ osp}(2|2)$ \cite{Alvarez:2011gd} .
Once expressed in terms of the same notations, the  super CS Lagrangian $\mathcal{L}(\mathbb{A})$ does coincide, modulo an overall rescaling, with the lagrangian (\ref{supercs}) {for the choice $\varepsilon=-1$}, if one identifies the spin-connection $\omega^{ij}$ 
in \eqref{A} with  $\omega_{(\varepsilon)}^i=\omega_{(-)}^i$ in \eqref{omegavareps}. The precise matching of the conventions and notations can be found in the Appendix A of \cite{Andrianopoli2018}.

The peculiar feature in \cite{Alvarez:2011gd}, which makes the model \commas unconventional", is that the spinor 1-form associated with the odd generator of the superalgebra is not a genuine $\mathrm{SO}(2)$-valued spin-3/2 field, being instead given in terms of a doublet of spin $1/2$ fields.
The defining equation, named by its authors \emph{matter Ansatz}, is:
\begin{equation}\label{gravchi}
  \psi_{\mu A}=i \,e^i_\mu\,\gamma_i\chi_A\,,
\end{equation}
where $\chi_A$ is a doublet of Majorana spinors and $e^i_\mu$ is interpreted as the dreibein of the $D=3$ space-time $\mathcal{M}_3$ where the CS Lagrangian is integrated on. It is intended as a background field, not a superfield and, as such, it is invariant under supersymmetry transformations.
{The resulting theory is then  naturally defined on the principal fiber bundle $[\mathcal{M}_3,\rm{OSp}(2|2)]$, under the peculiar assumption that the bosonic subgroup ${\mathrm{SO}(1,2)\subset\mathrm{OSp}(2|2)}$ of the fiber gauge group is identified with the Lorentz group in the tangent space of the base space-time $\mathcal{M}_3$ (also referred to  as \commas world-volume"). Indeed, this identification  is implicit  in the Ansatz (\ref{gravchi}), where the index $i$ of the $\gamma$-matrices is the same as the one for the world-volume dreibein $e^i_\mu$, and where it is assumed that the same set of $\gamma$-matrices will act both on the gravitino $\psi_{\mu A}^\alpha$, which belongs to the $\mathrm{OSp}(2|2)$ gauge connection, and on the world-volume spinor $\chi^\alpha_A$.

A crucial point of the construction is that torsion is allowed to be non-vanishing. It will be chosen, after exploiting all the symmetries of the theory (this will be discussed in detail later in this Section), of the form  \begin{align}\label{tor0}\mathcal{D}^{(\varepsilon)}e^i= \tau_0 \epsilon^i{}_{jk}e^j e^k,\end{align} with $\tau_0$ a constant.
When $\tau_0\neq 0$, the geometry of $\mathcal{M}_3$ can be traded, by changing the spin-connection into a torsionless one, for a locally AdS$_3$ one.

As shown in \cite{Alvarez:2011gd}, the unconventional realization of the CS gauge connection \eqref{A} via the Ansatz \eqref{gravchi}  does not spoil the main properties of the CS Action: under infinitesimal $\mathrm{OSp}(2|2)$ transformations, the modified connection  still transforms as expected for a gauge
connection as far as  bosonic gauge transformations are considered,  with the Action changing by a surface term, while supersymmetry is realized as a global symmetry (the details can be found in \cite{Alvarez:2011gd,Guevara:2016rbl}).

Plugging the matter Ansatz \eqref{gravchi} in the lagrangian \eqref{supercs},  we obtain:
\begin{align}
 \mathcal{L}_{(\varepsilon)}\,=&
\frac \ell 2\left( \omega_{(\varepsilon)}^i d \omega_{(\varepsilon)|i} -\frac 13  \omega_{(\varepsilon)}^i\omega_{(\varepsilon)}^j\omega_{(\varepsilon)}^k\varepsilon_{ijk} \right)+\,2\varepsilon \,e_i\,\mathcal{D}^{(\varepsilon)}e^i\,\bar{\chi}_A\chi_A+ \nonumber\\&- 4\, i \,\,\varepsilon \,\bar \chi_A \slashed{\nabla}^{{(\varepsilon)}} \chi_A\, {\rm e}\,d^3x\,
-\frac{\varepsilon}{2\ell} A\,dA \,,\label{supercs2}
\end{align}
where  
${\rm e}\equiv {\rm det}(e_\mu{}^i)$, and we have defined:
 \begin{equation}
 \slashed{\nabla}^{{(\varepsilon)}} \chi_A\equiv \gamma^i\,{\nabla}^{{(\varepsilon)}}_i \chi_A=
 \slashed{\mathcal{D}}^{(\varepsilon)}\chi_A -\frac \varepsilon{2\ell}\,A_i \epsilon_{AB}\gamma^i\chi_B\,.
 \end{equation}
The  Euler-Lagrange equation  of the spinor $\chi_A$ is:
\begin{equation}
\slashed{\nabla}^{(\varepsilon)}\chi_A  +{3i } \,\tau_0 \chi_A  =0  \,.\label{diracchi}
\end{equation}
Eq. \eqref{diracchi}
can be rewritten in terms of the Dirac spinor $\chi\equiv \chi_1 + i \chi_2$, thus taking the form of the Dirac equation:
\begin{equation}
\slashed{\nabla}^{(\varepsilon)}\chi   +{3i } \,\tau_0 \chi   =0  \,,\label{diracchi1}
\end{equation}
whose mass is given by  the torsion parameter $\tau_0$.
For $\tau_0= -\frac\varepsilon\ell$, which is the value chosen from now on, eq. \eqref{diracchi} is compatible with \eqref{asymptoticpsi}, which  gives, using  the matter Ansatz (\ref{gravchi}):
\begin{align}
\nabla^{(\varepsilon)}_i \chi_A = i \frac \varepsilon\ell \gamma_i \chi_A\,.\label{dchi}
\end{align}
Eq. \eqref{dchi}  implies the condition
\begin{equation}\label{dxx}
  d(\bar\chi\chi)=0\, ,
\end{equation}
so that the scalar quantity $\bar\chi\chi$ is a free constant of the model. As we are going to see, this is a consequence of having chosen  the torsion tensor as in \eqref{tor0}. Let me anticipate that  this  is not the most general expression for the torsion in $D=3$ (meaning, with this, the expression to which the torsion can reduce upon using the symmetries of the theory). \footnote{Indeed, as shown in \cite{Andrianopoli:2023dfm}, torsion in D=3 admits a more general expression, referred to as \emph{Beltrami torsion}. In that case, eq. \eqref{dxx} does not hold anymore. The effects of the  matter Ansatz on a geometry with Beltrami torsion have not been explored yet.}

The Euler-Lagrange equations for $\omega^{i}_{(\varepsilon)}$ and $A$ following from the lagrangian (\ref{supercs2}) or, equivalently, by substituting \eqref{gravchi} in \eqref{asymptotic}, are given by
\begin{align}\label{3Deq}
\mathcal{R}^{i}_{(\varepsilon)}&=- \frac{\varepsilon}{\ell}\,\bar{\chi}_A \chi_A \epsilon^{ijk} e_j e_k\,,\quad
dA = i \epsilon_{AB} \bar{\chi}_{A} \gamma^k \chi_{B}  e^i e^j \epsilon_{ijk}\,.
\end{align}
The equation of motion of the $e^i$ field implies the vanishing of the energy momentum tensor $T_{ij}=0$ \cite{Guevara:2016rbl,Andrianopoli2018}.
Further details on this point, and on all the construction, can be found in \cite{Alvarez:2011gd,Andrianopoli2018}.

\subsubsection{The role of the torsion}\label{torsion}

The Lagrangian \eqref{supercs2} describes the dynamics of a massive spin $1/2$-field  in a curved, torsionful $3D$-background, its mass being proportional to the torsion.
It could be puzzling to having found a local dynamics, here,  our starting point being  the  CS theory in \eqref{acs}, which is topological. The fact is that the matter Ansatz \eqref{gravchi} turned a background gravitino field, $\psi_A=\psi_{\mu A}dx^\mu$, which does not carry any propagating local degrees of freedom, into a couple of fields, namely the (non-propagating) dreibein 1-form $e^i= e^i_\mu dx^\mu$, and  the \emph{propagating} massive Dirac spinor $\chi$, which indeed satisfies the massive Dirac equation \eqref{diracchi1}.

One could wonder why this is possible.
The reason can be traced back to the fact that the Ansatz \eqref{gravchi} introduces in the Lagrangian a dependence on the world-volume metric, just as a gauge-fixing does. \footnote{ An analysis of this issue as a covariant gauge-fixing of the 3D gauge-supersymmetry  can be found in \cite{Andrianopoli:2019sqe}, and an explicit application is given in
\cite{Andrianopoli:2021sdx}. However, the matter ansatz \eqref{gravchi}   allows for more general interpretations, as discussed in \cite{Valenzuela:2022gbk,Valenzuela:2023aoa}, depending on the choice to use, or not, the Dirac conjecture in the Hamiltonian analysis of the constraints. As recently pointed out in \cite{Francois:2024xqi}, this issue can be generally reformulated as a special case of \emph{dressing} \cite{Francois:2024laf}, within the systematic tool  named Dressing Field Method \cite{Francois:2024laf,Francois:2024rdm},  treating in a unified way the symmetric structure of general relativistic and gauge field theories.}
As a consequence, the  dreibein $e^i_\mu$ is included among  the fields of the Lagrangian. Moreover, the $\mathrm{SO}(1,2)\subset \mathrm{OSp}(2|2)_{(\varepsilon)}$  connection $\omega^{ij}_{(\varepsilon)}$ of the gauge group is  identified with the Lorentz spin connection acting on the dreibein $e^i$ of the base space $\mathcal{M}_3$. 

However, let us remark that implicit in the matter Ansatz \eqref{gravchi}, as observed in \cite{Alvarez:2011gd,Guevara:2016rbl,Andrianopoli2018},  is the local scale invariance under the so-called Nieh-Yan-Weyl (NYW) symmetry \cite{Nieh:1981xk,Chandia:1997hu,Hughes:2012vg,Parrikar:2014usa}
\begin{align}\label{NWsymm0}
e^i\to\lambda(x)\,e^i\,, \quad\;\; \chi_A\to\frac{1}{\lambda(x)}\,\chi_A\;, \qquad \lambda\neq 0\,,
\end{align}
which leaves the gravitino, and the whole theory, invariant.
Then, the normalization of the spinor $\chi_A$ is immaterial, as it amounts to choosing a specific metric, with dreibein $e^i$, in the conformal class of metrics related by $\mathrm{SO}(1,1)$-rescalings $\lambda(x)$.
Related to this, let me remark that the mass  of the spinor $\chi$ is proportional to the fully antisymmetric   component of the space-time torsion 2-form $\hat T^i \equiv \mathcal{D}^{(\varepsilon)}e^i$, in the decomposition of the torsion tensor into irreducible $SO(1,2)$ representations \cite{Alvarez:2011gd,Andrianopoli2018,Andrianopoli:2023dfm}:
{\begin{align}
\label{torcomp}\mathcal{D}^{(\varepsilon)}e^i = \hat T^i{}_{jk} e^j\wedge e^k= \beta \wedge e^i + \tau e^j \wedge e^k \epsilon^i{}_{jk} + \mathring{T}^{i\ell}\epsilon_{jk\ell}e^j \wedge e^k
\end{align}
where $\beta$ is a 1-form, $\tau$ a scalar and $\mathring{T}^{i\ell}$ a traceless tensor, satisfying $\mathring{T}^{i\ell} g_{i\ell}=0$. \footnote{In the following, I will set the component $\mathring{T}^{ij}=0$. Some words on its possible relevance will be given in the concluding Section of the present contribution.
}
Consistency of torsion with the Lorentz spin-structure implies the Bianchi identity $(\mathcal{D}^{(\varepsilon)})^2 e^i=\mathcal{R}^{ij}_{(\varepsilon)}\wedge e_j$.
In the case at hand, plugging  the condition \eqref{gravchi} in the first of \eqref{asymptotic} gives:
\begin{align}
   \label{rosp22}\mathcal{R}^{i}_{(\varepsilon)} = \frac \varepsilon\ell \bar\chi_A \chi_A e^j\wedge e^k\epsilon_{ijk}= \frac 12 \epsilon^{i}{}_ {jk}\mathcal{R}^{jk}_{(\varepsilon)}\,,
\end{align}
which implies $(\mathcal{D}^{(\varepsilon)})^2 e^i=0$.
This admits the solution
\begin{align}
    \beta = -\frac{d\tau}\tau\,.\label{betatri}
\end{align}

Under a Weyl rescaling of the dreibein:
\begin{subequations}
\begin{align}
    e^i &\to \lambda(x) e^i\quad    \Rightarrow\quad \mathcal{D}^{(\varepsilon)}e^i \to \lambda\left(\mathcal{D}^{(\varepsilon)}e^i+\frac {d\lambda}\lambda\wedge e^i
    \right)\,,
\end{align}
\end{subequations}
 the torsion components rescale as follows \cite{Alvarez:2011gd,Andrianopoli:2019sip,Andrianopoli:2023dfm}:
\begin{align}
 \beta \to  \beta   +\frac {d\lambda}\lambda\,,\quad \tau \to  \frac {\tau}\lambda 
 ,.\label{rescal}
\end{align}

Eq.s \eqref{betatri} and \eqref{rescal} imply that, when the torsion satisfying \eqref{betatri} is non-vanishing,  it is always possible to dispose of the NYW symmetry \eqref{NWsymm0} to rescale $\tau$ to a constant value $\tau_0$ and, correspondingly, $\beta$ to zero.
In this case, changing the spin-connection $\omega^i_{(   \varepsilon)}$ to a torsionless one, $\omega'^i$:
\begin{align}
 \omega'^i=\omega^i_{(   \varepsilon)} +\tau_0 e^i
\end{align}
gives for the   Riemann tensor, instead of \eqref{rosp22}:
\begin{align}
\mathcal{R}^{ij}[\omega']= \left(\tau_0^2+\frac \varepsilon\ell \bar\chi_A \chi_A \right)e^i\wedge e^j\,.
\end{align}
As is well known, this corresponds to change the space-time geometry  into a locally AdS$_3$ one, that is to trade the constant torsion-component $\tau_0$ for a negative cosmological constant with AdS radius $\ell' = \frac 1{|\tau_0|}$.
 This implies that, for $\tau_0\neq 0$, the manifest symmetry of the base space $\mathcal{M}_3$ is the AdS$_3$ isometry group $\mathrm{SO}(2,2)$.
Having identified the spin-bundle on the base manifold $\mathcal{M}_3$ with the gauge Lorentz group of the connection \eqref{A},
it is then natural to embed the $\mathrm{OSp}(2|2)$ model originally considered in \cite{Alvarez:2011gd} into the $\mathrm{OSp}(2|2)_{(\varepsilon)}\times \mathrm{SO}(1,2)$ Chern-Simons theory belonging to the family of $\mathcal{N}=2$ A-T supergravities, which was the object of Section \ref{ATsugra} (and identify $\ell'=\ell$).
As we will see later, this is precisely the framework where a possible analog description in terms of a graphene-like system emerges.

Let me conclude by remarking that the condition \eqref{betatri} is not the most general solution to the torsion Bianchi identity in $D=3$ Einstein backgrounds. Indeed, as found in \cite{Andrianopoli:2023dfm}, in this case the general solution of the torsion Bianchi identity for the torsion \eqref{torcomp}, at $\mathring{T}^{ij}=0$, allows for a more general $\beta$ term which is not an exact form as in \eqref{betatri}, but satisfies instead a differential equation as a   self-dual gauge field in odd dimensions \cite{Townsend:1983xs}:
\begin{align}
    {}^*d\beta= -2\tau\left(\beta +\frac{d\tau}\tau\right)
\end{align}
with $\tau$ behaving as its mass. It is  analog to the equation for a  chiral Beltrami fluid, and for this reason it was named \commas Beltrami-torsion".

\subsection{Relation with 4D Supergravity}\label{backto4}

As discussed above, the massive spin-$1/2$ model of \cite{Alvarez:2011gd}, which was the object of Section \ref{ususy}, propagates on a world-volume which, for $\tau_0\neq 0$, can be expressed as an AdS$_3$ background. As such, as we discussed above, it is naturally embedded in
A-T supergravity with supergroup $\mathrm{OSp}(2|2)_{(\varepsilon)}\times \mathrm{SO}(1,2)$,  presented in Section \ref{ATsugra} and showed to be able to reproduce,  at the Lagrangian level,
 a specific holographic limit of AdS$_4$ supergravity, as discussed in Section \ref{aads4}. Further details can be found in \cite{Andrianopoli2018}.

Let us now discuss more in detail, following \cite{Andrianopoli2018}, what we gain from obtaining the model of \cite{Alvarez:2011gd} at the boundary of AdS$_4$ supergravity.
Two interesting issues emerge in the comparison and are worth to be discussed:
\par One consequence is purely $3D$,  and amounts to the fact that we have two inequivalent dreibein,  namely the space-time dreibein $e^i$ introduced in \eqref{gravchi}, which is supersymmetry-invariant and torsionful (with respect to the spin connection $\omega^i_{(\varepsilon)}$).
The other one is the superfield dreibein $E^i$ of A-T supergravity, which is related to the bosonic connections $\omega^{ij}_{(\pm \varepsilon)}$ via \eqref{omegavareps}, and  whose torsion (with respect to the supertorsionless spin-connection $\omega^i=\frac 12(\omega^i_{(\varepsilon)}+\omega^i_{(-\varepsilon)})$) is given  in (\ref{asymptoticeq}).
  The presence of two sets of  dreibein, $e^i$ and $E^i$, 
  seems at first sight a puzzling feature of our model.
To clarify this point, let us explicitly compare the torsion equation for the  bosonic dreibein $e^i$, eq. \eqref{tor0}, with $\tau_0=-\frac\varepsilon \ell $:
\begin{align}\label{tor1}\mathcal{D}_{(\varepsilon)}e^i= -\frac\varepsilon \ell \epsilon^i{}_{jk}e^j e^k,\end{align}with the super-torsion equation in (\ref{asymptoticeq}) for the super-dreibein $E^i$,  after using the matter Ansatz  \eqref{gravchi}.
If  written  in terms of the same spin connection $\omega^i_{(\varepsilon)}$, the second 
equation in (\ref{asymptoticeq}) reads:
\begin{eqnarray}
\mathcal{D}_{(\varepsilon)} E^i&=&-\frac{\varepsilon}{\ell}\,\epsilon^{ijk}\,E_j\, E_k + \frac 12 \bar\chi_A \chi_A \epsilon^{ijk} e_j \,e_k\,.\label{d+E}
\end{eqnarray}
Eq. (\ref{d+E}) contains in its right-hand side both sets of dreibein, hinting at a relation expressing the $E^i$ in terms of the space-time dreibein $e^i$ and of the spinor $\chi_A$.
Consistency of the two descriptions determines  $E^i$ to be:
\begin{align}\label{Ee}
E^i &=  M(\bar\chi\chi) \,e^i\,,\quad \text{with:  } M(\bar\chi\chi)=\left(1 +\varepsilon \,\frac\ell 2\,  \bar{\chi}\chi -\frac{\ell^2}4 (\bar{\chi}\chi)^2\right)\,.
\end{align}
The details of the calculation can be found in \cite{Andrianopoli2018}.
\par The second issue emerging from the interpretation of the model of \cite{Alvarez:2011gd} in terms  of AdS$_4$ supergravity fields, in the special limit discussed in Section \ref{limit}, regards the meaning of  the spinor $\chi$.
As defined in \eqref{gravchi}, $\chi_A$ is the spin-$1/2$ projection of the $D=3$ gravitino $\psi_A$:
\begin{equation}\label{chi}
\chi_A= -\frac i 3 \,\gamma_i\, e^{i|\mu}\psi_{A\,\mu} \,.
\end{equation}
On the other hand, the ${\mathcal N}=2$ gravitino in $D=4$, $\Psi_{A\,\hat\mu}$, where $\hat{\mu}
=0,1,2,3$, satisfies the condition
\begin{equation}\label{gra4}
  \Gamma_a V^{a|\hat{\mu}}\Psi_{A\,\hat\mu}=
  0\,,
\end{equation}
which projects out its spin-$1/2$ component all over the bulk of  $D=4$ superspace.
Close to the boundary at $r\rightarrow \infty$, expanding Eq. (\ref{gra4}) in powers of $\ell/r$ as in \ref{limit} and using the boundary behavior (\ref{beha}), eq. (\ref{gra4}) gives, for the radial component $\Psi_{A\,3}$ of the gravitino fields (decomposed in chiral projections with respect to $\Gamma^3$):
\begin{equation}\label{psir}
\Psi_{+\,A\,3}=\frac{3i \varepsilon}{2M(\bar \chi \chi)}\,\left(\frac{\ell}{r}\right)^{\frac{5}{2}}\,\begin{pmatrix}
\chi_A\\{\bf 0}\cr\end{pmatrix}\,\,,\,\,\,\,\,\Psi_{-\,A\,3}=\frac{3i }{M(\bar \chi \chi)}\,\left(\frac{\ell}{r}\right)^{\frac{3}{2}}\,\begin{pmatrix}
 {\bf 0}\\\chi_A\cr
\end{pmatrix}\,.
\end{equation}
 In the comparison, we gain the possible interpretation of the spinor $\chi_A$ that, as will be discussed in the following section, can be associated with the pseudo-particle describing the electronic properties of graphene-like materials near the Dirac points, as originating from the radial component of the gravitino field of $D=4$ supergravity.

\section{Unconventional Supersymmetry and \\Graphene-like Systems} \label{graf}
In the present Section, following \cite{Andrianopoli:2019sip}, we briefly discuss how  unconventional supersymmetry can provide an  effective description of the electronic properties of graphene-like materials.
This results in a top-down approach to understanding the possible origin of the supersymmetric phenomenology of these physical systems \cite{Ezawa:2006cr,Dartora:2013psa}.
Since the original model of \cite{Alvarez:2011gd} does not feature enough symmetries to accommodate the quantum numbers of the electrons wave-function of this kind of systems, it is necessary to extend the analysis presented in Section \ref{3d}  to a more general element of the A-T family of $\mathcal{N}=(p+q)$-extended supergravities ${\rm OSp}(p|2)_+\times {\rm OSp}(q|2)_-$.
 Applying the matter Ansatz \eqref{gravchi} to this more general case, an effective model for the massive spin-$1/2$ fields $\chi_A$ (with now $A= 1,\cdots ,\mathcal{N}$) on a curved background in the presence of  a larger internal symmetry group ${\rm SO}(p)\times {\rm SO}(q)$ is obtained. This allows to introduce extra internal degrees of freedom which make the model suitable for the application to the description of graphene.

 The supersymmetry of the boundary model is defined by the partition $(p,q)$ of $\mathcal{N}$.
In this more general case, the fermionic fields $\chi_A$ naturally split into two sets, $\chi_{a_1}$ and $\chi_{a_2\,}$, with $a_1=1,\dots, p$ and $a_2=1,\dots, q$, which correspond to the representations ${\bf \left(\frac{1}{2},0\right)}$ and ${\bf \left(0,\frac{1}{2}\right)}$ of the AdS$_3$ isometry group $\mathrm{SL}(2,\mathbb{R})_+\times \mathrm{SL}(2,\mathbb{R})_-$.
For even $p$ and $q$, both sets of spinors satisfy  massive Dirac equations, with masses depending on the torsion of the $D=3$ spacetime as in the original model \cite{Alvarez:2011gd}.

I will focus here in particular on the special case $p=q=2$, where a manifest \emph{reflection symmetry} emerges in the model, under which the fermions in the two sets are interchanged, and where the description is naturally formulated in terms of Dirac spinors. In this case, as I will review, they are suitable to describe the wave functions of the $\pi$-electrons in graphene at the two inequivalent Dirac points ${\bf K},\,{\bf K}'$. The details can be found in \cite{Andrianopoli:2019sip}.

\subsection{AdS$_4$ $\mathcal{N}$-extended supergravity and its boundary A-T Supergravity}

This section aims to extend, following \cite{Andrianopoli:2019sip}, the discussion of Section \ref{aads4} on the asymptotic behavior of AdS$_4$ supergravity in the presence of nontrivial boundary conditions, from $\mathcal{N}=2$ to a general $\mathcal{N}$-extended theory.
From the analysis of Section \ref{aads4} we learned that, to preserve asymptotically, at radial infinity, the full supersymmetry, the set of super-fieldstrengths should tend to the maximally  supersymmetric AdS$_4$ vacuum configuration with global symmetry $\mathrm{OSp}(\mathcal{N}|4)$, corresponding to the superconformal symmetry of the boundary theory.
  This is realized if  all the scalars and spin-$1/2$ fields are frozen at the conformal boundary to their vacuum values,   the remaining fields then satisfying at the boundary the $\mathfrak{osp}(\mathcal{N}|4)$ Maurer-Cartan equations.
  In this case, the boundary contributions to the Lagrangian will not depend on the scalars and spin-$1/2$ fields of the $\mathcal{N}$-extended AdS$_4$-supergravity, making the analysis strictly parallel to the one for pure $\mathcal{N}=2$.
  Close to the boundary, supergravity effects correspond to fluctuations of the fields  about the  $\mathrm{OSp}(\mathcal{N}|4)$ background.
Under the above conditions, let us then start with the asymptotic  $r\to \infty$ limit of the theory, given in terms of the Maurer-Cartan equations of the $\mathfrak{osp}(\mathcal{N}|4)$ algebra.
It is written in terms of the $\mathrm{SO}(1,3)$ spin-connection $\omega^{ab}$, of the bosonic vielbein $V^a$ ($a=0,1,2,3$), of the Majorana-spinor gravitino 1-form $\Psi^A_\alpha$, and of the $\mathrm{SO}(\mathcal{N})$ gauge connection $A^{AB}$  (with $A,B,...=1,\cdots ,\mathcal{N}$) \footnote{Note that the normalization of the gauge connection here is  the natural generalization of the one of  Section \ref{aads4}, but is different from the one in \cite{Andrianopoli:2019sip}. Moreover, we recall that, being $A,B...$ fundamental indices  of  $\mathrm{SO}(\mathcal N)$, we will not distinguish  their upper or lower position.}:
\begin{equation}
\begin{split}
&d\omega^{ab}+\omega^{a}{}_{c}\wedge\omega^{cb}-\ell^{-2}V^a\wedge V^b-\frac{1}{2{\ell}}\left(\overline\Psi_A\wedge\Gamma^{ab}\Psi_{A}\right)=0, \\
&dV^a+\omega^{a}{}_{b}\wedge V^b-\frac{i}{2}\left(\overline\Psi_A\wedge\Gamma^a\Psi_A\right)=0, \\
&dA^{CD}+A^C{}_B\wedge A^{BD}-\left(\overline\Psi{}^C\wedge\Psi^D\right)=0, \\
&d\Psi^A+\frac{1}{4}\,
\omega^{ab}\wedge\Gamma_{ab}\Psi^A+\frac{i}{2{\ell}}\,
V^a\wedge\Gamma_a\Psi^A-\frac 1{2\ell}A^{AB}\wedge
\Psi^B=0.\label{MCADS}
\end{split}
\end{equation}
The above relations generalize eq.s \eqref{lagsuperN2} to the $\mathcal{N}$-extended case, and are the conditions satisfied  asymptotically by the  super field-strengths of $\mathcal{N}$-extended AdS$_4$ supergravity to preserve the full amount of supersymmetry at the boundary.

Then, we can proceed on the same lines as in Section \ref{aads4}, and rewrite the Maurer-Cartan equations in a form which is covariant with respect to the Lorentz group at the spatial boundary. This is achieved by splitting the  index $a$ into $a=(i,3)$, and making explicit the $\mathrm{SO}(1,1)$-grading of the fields, by defining the 1-forms:
$E^i_\pm \equiv \pm\frac{1}{2}\left(V^i\mp \ell\omega^{3i}\right)$
and by decomposing the gravitini 1-forms in their \commas chiral" components with respect to the matrix $\Gamma^3$:
$
\Psi^{A}=\Psi^A_++\Psi^A_-$,
 $\Gamma^3\Psi^A_\pm=\pm i\,\Psi^A_\pm$.
With these definitions,  eq.s \eqref{MCADS} become
\begin{subequations} \label{eq:boundary}
\begin{align}
&\left[d\omega^{ij}+\omega^{i}{}_{k}\wedge\omega^{kj}+\frac{4}{{\ell}^2}\,E_+^{[i}\wedge E_-^{j]}-\frac{1}{{\ell}}\left(\overline\Psi^A_+\wedge\Gamma^{ij}\Psi_{A-}\right)\right]_{\partial \mathcal{M}}\!=0\,,
\\
&\left[dE_\pm^i+\omega^i{}_j\wedge E_\pm^j\mp\frac{1}{{\ell}}\,E^i_\pm\wedge V^3\mp\frac{i}{2}\left(\overline\Psi^A_\pm\wedge\Gamma^i\Psi_{A\,\pm}\right)\right]_{\partial \mathcal{M}} \!=0\,,
\\
&\left[dV^3-\frac{1}{\ell}\,(E^i_+ + E^i_-)\wedge V_i+\overline\Psi^A_-\wedge\Psi_{A+}\right]_{\partial \mathcal{M}}\!=0\,,
\\
&\left[dA^{CD}+A^C{}_M\wedge A^{MD}-2\left(\overline\Psi{}^{[C}_+\wedge\Psi^{D]}_-\right)\right]_{\partial \mathcal{M}}\!=0\,,
\\
&\left[d\Psi^{M\beta}_\pm+\frac{1}{4}\,\omega^{ij}\wedge\left(\Gamma_{ij}\Psi^M_\pm\right)^\beta\pm
\frac{i}{{\ell}}\,E_\pm^i\wedge\left(\Gamma_i\Psi^{M}_\mp\right)^\beta \pm\frac{1}{2{\ell}}\,V^3\wedge\Psi^{M\beta}_\pm +\right.\nonumber\\
&\left. \quad -\frac 1{2\ell}\delta^M{}_{[C}\delta_{D]B}\,A^{CD}\wedge\Psi^{B\beta}_\pm\right]_{\partial \mathcal{M}}\!=0\,.
\end{align}
\end{subequations}
Now, exactly as in the $\mathcal{N}=2$ case,  the actual boundary  theory can be obtained   by suitably choosing the explicit expansion of the fields, close to the asymptotic boundary.
For an asymptotic    locally  AdS$_3$ geometry, we should impose the same boundary behavior as in \eqref{beha}. More precisely, for the  $\mathrm{SO}(2,3)$ fields $E^i_\pm, \omega^{ij}, V^3$, the normalizations of the boundary fields  are chosen exactly as in \eqref{beha}, and
for the $\mathrm{SO}(\mathcal{N})$ gauge connection we require:
    \begin{align}
A^{AB}(r,x)= -2\ell\,A^{AB}_\mu(x)\,dx^\mu+\mathcal{O}\left(\frac{{\ell}}{r}\right)\,.
\end{align}
For the spinors, to choose a boundary behavior compatible with all the $(p,q)$ partitions of the family of A-T $\mathcal{N}=(p+q)$-extended supergravities, we impose:
\begin{align}
\Psi^A_{+\,\mu}(r,x)\,dx^\mu &=
\sqrt\frac{r}{2\ell}\begin{pmatrix}
  \psi^A(x)  \\
  \bf 0
\end{pmatrix}+O\left(\frac{{\ell}}{r}\right), \\
\Psi^A_{-\mu}(r,x)\,dx^\mu &=
\sqrt\frac{{\ell}}{2 r}\begin{pmatrix}
\bf 0  \\
  \eta^{AB}\psi^B(x)
\end{pmatrix}+O\left(\frac{{\ell}}{r}\right),
\end{align}
where $\eta_{AB}$ is a symmetric metric
 breaking  the $D=4$ R-symmetry group: ${{\rm O}(\mathcal{N})\rightarrow {\rm O}(p)\times {\rm O}(q)}$, $p+q=\mathcal{N}$ (so that  the index $A=1,\dots, \mathcal{N}$  splits into $A=(a_1,a_2)$, where $a_1=1\,\dots,\,p $ and $a_2=1\,\dots,\,q$):
\begin{align}\label{etadiag}
\eta_{AB}=\begin{pmatrix}
\delta_{a_1 b_1} & \textbf{0} \\
\textbf{0} & -\delta_{a_2 b_2}
\end{pmatrix}\,.
\end{align}
Using the above boundary conditions in the Maurer-Cartan equations \eqref{eq:boundary}, at leading order in $\ell /r$ they reduce to:
\begin{subequations}\label{AT-likecurv}
\begin{align}
&R_\pm {}^i\equiv d\Omega_\pm^i-\frac{1}{2}\epsilon^{ijk}\Omega_{\pm \,j} \wedge\Omega_{\pm\,k}=\pm\frac{i}{{\ell}}\left(\overline\psi_\pm\wedge\gamma^{i}\psi_\pm\right)\,;
\\[\jot]
&\mathcal{D}[\Omega_\pm,\,A_\pm]\psi_\pm=0\,,\label{eq:DOmAPsi}
\\[\jot]
&\mathcal{F}^{a_1 b_1}\equiv dA^{a_1b_1}+A^{a_1}{}_{c_1}\wedge A^{c_1b_1}=-\frac{1}{{\ell}}\,\left(\overline\psi^{a_1}\wedge\psi^{b_1}\right)\,,
\\
&\mathcal{F}^{a_2 b_2}\equiv
dA^{a_2b_2}+A^{a_2}{}_{c_2}\wedge A^{c_2b_2}=\frac{1}{{\ell}}\left(\overline\psi^{a_2}\wedge\psi^{b_2}\right)\,,
\end{align}
\end{subequations}
where we have introduced the compact definitions:
\begin{equation}
\begin{split}
&\Omega_{(\pm)}^i\equiv\,\omega^i\pm\frac{E^i}{{\ell}}\,,\quad
A_+\equiv(A^{a_1b_1})\;,\quad
A_-\equiv (A^{a_2b_2})\;,\\
&\psi_+\equiv(\psi^{a_1})\,,\quad\mathcal{D}[\Omega_+,\,A_+]\psi_+ \equiv \left(d\psi^{a_1}+\frac{i}{2}\Omega^i_+\wedge\gamma_i\psi^{a_1}+
A^{a_1b_1}\wedge\psi^\beta_{b_1}\right)\;,\\
&\psi_-\equiv(\psi^{a_2})\,,\quad\mathcal{D}[\Omega_-,\,A_-]\psi_- \equiv \left(d\psi^{a_2}+\frac{i}{2}\Omega^i_-\wedge\gamma_i\psi^{a_2}+
A^{a_2b_2}\wedge\psi_{b_2}\right)\;.
\end{split}
\end{equation}%
Eq.s \eqref{AT-likecurv} are the Maurer-Cartan equations of the  $\mathfrak{osp}(p|2)_+\oplus\mathfrak{osp}(q|2)_-$  superalgebra, and can be derived as Euler-Lagrange equations from the A-T supergravity Lagrangian 3-form \cite{Achucarro:1987vz}
\begin{align}\label{Lagrangianpm}
\mathcal{L}=\;&\mathcal{L}_{(+)}-\mathcal{L}_{(-)}-\frac{1}{2}\,d(\Omega_+{}_k\wedge \Omega^k_-)\,,
\end{align}
where
\begin{align}
\mathcal{L}_{(\pm)}\equiv \;&\frac{1}{2}\,\left(\Omega_{\pm\,i}d\Omega_\pm^i-\frac{1}{3}\,
\epsilon_{ijk}\,\Omega_\pm^i\wedge \Omega_\pm^j\wedge \Omega_\pm^k \right)\pm \frac{2}{\ell}
\overline{\psi}_\pm\wedge\mathcal{D}[\Omega_\pm,\,A_\pm]\psi_\pm+\nonumber\\
&+\,{\rm Tr}\left(A_\pm\wedge dA_\pm+\frac{2}{3}\,A_\pm\wedge A_\pm\wedge A_\pm\right)\,.
\end{align}

\subsubsection{Reflection transformations and the symmetric case {$p=q$}  }\label{parity}
It is useful to consider the effect of a spatial parity transformation, say in the Y-axis, on the 3-dimensional  A-T model above. This transformation can be characterized as a spatial reflection: ($t\rightarrow t,\; x\rightarrow -x,\; y\rightarrow y$),  implemented  by the matrix $\mathcal{O}_Y={\rm diag}(+1,-1,+1)$, under which:
\begin{align}
E^i\rightarrow \mathcal{O}_Y{}^{i}{}_{j}\,E^j\,,
\quad
\omega^i\rightarrow -\mathcal{O}_Y{}^{i}{}_{j}\,\omega^j\,\quad \Rightarrow\quad \Omega^i_\pm\rightarrow -\mathcal{O}_Y{}^{i}{}_{j}\,\Omega^j_\mp\,.\label{parity1}
\end{align}
As a consequence, the parity transformation acts on the $SO(2,2)$ subgroup of the A-T supergroup
by interchanging the $+$ and $-$ sectors: $$\mathrm{SO}(1,2)_+\times \mathrm{SO}(1,2)_-\quad \longrightarrow \quad \mathrm{SO}(1,2)_-\times \mathrm{SO}(1,2)_+\,.$$
In the special case $p=q$, this discrete  transformation is an invariance of the A-T theory. To extend the invariance to the full supergroup, we have to extend the action of parity to the fermionic and gauge sectors as well.
This can be obtained by imposing (see \cite{Andrianopoli:2019sip} for the details):
\begin{align}
\psi_\pm\rightarrow \tilde{\psi}_\pm=\sigma^1\,\psi_\mp\,,
\quad\;\;
A_\pm\rightarrow \tilde{A}_\pm=A_\mp\,.
\label{parity2}
\end{align}
 Then, under the parity transformation, in this case 
 $\mathcal{L}$ in \eqref{Lagrangianpm} is odd. 


\subsection{Imposing the matter Ansatz}
We are now ready to discuss the consequences of imposing the matter Ansatz \eqref{gravchi}:
\begin{equation}\label{gravchiext}
  \psi_{\mu A}=i \,e^i_\mu\,\gamma_i\chi_A\,,
\end{equation}
as for the classical model of \cite{Alvarez:2011gd,Andrianopoli2018}, in the more general case of the gauge group ${{\rm OSp}(p|2)_+\times {\rm OSp}(q|2)_-}$, where now $A=(a_1,a_2)= 1,\cdots ,\mathcal{N}$. We will let for the moment $p$ and $q$ be generic, but we will then explicitly consider only the case $p=q=2$. As for the $\mathcal{N}=2$ case, implicit in the Ansatz (\ref{gravchi}) is the NYW invariance, eq. \eqref{NWsymm0}, under local rescaling of $e^i$ and $\chi_A$,
which leaves the gravitino, and the whole theory, invariant.

Implementing the Ansatz (\ref{gravchi}) in the ${\rm OSp}(p|2)_+\times {\rm OSp}(q|2)_-$ structure equations (\ref{AT-likecurv}) for the bosonic curvatures, yields
\begin{subequations}
\label{bosonicAVZ}
 \begin{align}
R^i_\pm&=\pm \frac{1}{{\ell}}\,\bar{\chi}_\pm \chi_\pm\epsilon^{ijk}\,e_j \wedge e_k\,,\label{r1}
\\
\mathcal{D}[\Omega_\pm] E^i&=\mp \frac{1}{{\ell}}\,\epsilon^{ijk}\,E_j \wedge E_k+\frac{1}{2}\,(\bar{\chi}_+\chi_++\bar{\chi}_-\chi_-)\,\epsilon^{ijk}\,e_j \wedge e_k\,,
\\
\mathcal{F}^{a_1 b_1}&=-\frac{i}{{\ell}}\,\left(\bar\chi^{a_1}\gamma^i\chi^{b_1}\right)\,\epsilon_{ijk} e^j\wedge e^k\,,\\
\mathcal{F}^{a_2 b_2}&=\frac{i}{{\ell}}\,\left(\bar\chi^{a_2}\gamma^i\chi^{b_2}\right)\,\epsilon_{ijk} e^j\wedge e^k\,,
\end{align}
\end{subequations}
where $\chi_+\equiv (\chi_{a_1})$\, and \,$\chi_-\equiv (\chi_{a_2})$.
As already observed in Section \ref{backto4}, the Ansatz \eqref{gravchiext} introduces a redundant set of drei\-bein:  the (non-supersym\-metric)  space-time dreibein, $e^i$,  of the world-volume $\mathcal{M}_3$ where the CS Action is integrated on, and the supergravity bosonic vielbein $E^i$, which is a part of the CS gauge connection. Still,  the Lorentz groups on the base space and on the gauge fiber get identified by the Ansatz.}
The redundancy will be fixed in the following, by postulating the relation
\begin{align}\label{relatingvielbein}
E^i=f\,e^i\,,
\end{align}
where $f$ is some indeterminate function%
\footnote{{Here we assume, as in Section \ref{backto4},  the 1-form $E^i$ to be parallel to $e^i$. More general situations may occur, with interesting physical implications. I will comment on this point in the final lines of the contribution. }}, and then
requiring consistency of the construction, on the same lines as in Section \ref{backto4} (see eq. \eqref{Ee}).

The torsion field plays a relevant role here, because, as already observed in the simpler case discussed in Section \eqref{ususy}, after plugging the matter Ansatz \eqref{gravchiext} in the Lagrangian \eqref{Lagrangianpm}, the torsion of the dreibein $e^i$ will enter the Lagrangian as a mass term for the Dirac spinor.

In the $p=q=2$ case, where  both  the $+$ and $-$ parts of the A-T superalgebra have an odd sector, thus contributing both to the set of spinors in the resulting model,  we have several possible candidates as the torsionful  spin connection in the covariant derivative of $e^i$. Two natural choices are  $\Omega^i_\pm$, both corresponding to flat super-torsionful spin-connections on the fiber (in the 2 sectors $\pm$ of A-T supergravity).
These two possible choices of spin-connection lead to inequivalent decompositions of the torsion in components, that is  (setting to zero, as before, the symmetric traceless components):
\begin{align}\label{integrability}
T^i_{\pm}=\mathcal{D}[\Omega_\pm]e^i=\beta_\pm e^i+\tau_\pm\epsilon^{ijk} e_j\wedge e_k\,,
\end{align}
where $\beta_\pm$ and $\tau_\pm$ are 1- and 0-forms, respectively.
Under the NYW symmetry transformation (\ref{NWsymm0}), the above expressions retain their form provided $\beta_\pm$ and $\tau_\pm$ change according to
\begin{equation}\label{NYWbeta}
\beta_\pm\rightarrow \beta_\pm +\frac{d\lambda}{\lambda}\,,
\qquad
\tau_\pm\rightarrow \frac{1}{\lambda}\,\tau_\pm \,.
\end{equation}
The integrability condition on (\ref{integrability}) yields, using (\ref{r1})
$$\mathcal{D}[\Omega_\pm]^2e^i=-\epsilon^{ijk}\,R_{\pm\,j}\,e_k=0\,,$$
 which admits the solution (for $\tau_\pm\neq 0$)\begin{equation} \label{Beta=dlog}
\beta_\pm=-\frac{d\tau_\pm}{\tau_\pm}=-d{\rm ln}(|\tau_\pm|)\,.
\end{equation}
On the other hand, we could instead consider the spin connection $\omega^i$ in \begin{align}
 \Omega^i_\pm = \omega^i \pm E^i/\ell\label{opm}   \,,
\end{align} which is supertorsionless (as it can be derived from  \eqref{AT-likecurv}) and then torsionful, and  compute the corresponding covariant derivative of $e^i$ by defining
\begin{equation}\label{Dome}
 \mathcal{D}[\omega]e^i=\beta \wedge e^i+\tau\,\epsilon^{ijk} e_j\wedge e_k\:.
 \end{equation}
At this point, we have introduced several unknown functions/parameters ($f$, $\beta_\pm$, $\tau_\pm$, $\beta$, $\tau$) and we are ready to exploit all the symmetries of the model to constrain them, and then interpret the result in  physical terms. I will sketch here the main points, referring to \cite{Andrianopoli:2019sip} for a more general and comprehensive discussion.
From the definition \eqref{opm}, assuming (\ref{relatingvielbein}), the covariant derivative of $e^i$ can also be written as \begin{equation}\mathcal{D}[\Omega_\pm]\,e^i= \mathcal{D}[\omega]\,e^i {\mp} (f/\ell)\, \epsilon^{ijk}e_j\wedge e_k\,,\end{equation} from which one obtains
\begin{equation}
 (\beta_+ - \beta_-)\,e^i +(\tau_+ -\tau_-{+}2f/\ell)\,\epsilon^{ijk}e_j \wedge e_k =0 .\label{differenceseq}
 \end{equation}
Comparing equation (\ref{Dome}) with  (\ref{differenceseq}), one finds
\begin{equation}\label{betatau}
\beta_+=\,\beta_-=\,\beta\,, \qquad \tau_+ +\, \frac{f}{\ell}\,=\,\tau_- -\, \frac{f}{\ell}\,=\,\tau\,.
\end{equation}
Note that, in the absence of global obstructions, the 1-form $\beta$ can be disposed of through a NYW transformation (\ref{NYWbeta}) \cite{Alvarez:2011gd} setting it to zero and correspondingly setting $\tau$ to a constant \footnote{As already observed in Section \ref{3d}, we are disregarding here a more general possible solution to the torsion Bianchi identity in 3D, corresponding to the so-called chiral Beltrami-torsion, of which we were not aware at the moment of the publication \cite{Andrianopoli:2019sip}. This more general   torsion was investigated later in \cite{Andrianopoli:2023dfm,Andrianopoli:2024twc}.} .
This would leave the residual invariance  under  rigid NYW transformations, which can in turn be used to fix the value of either $\tau_+$ or $\tau_-$ at will since, for $\beta=0$, $\tau_\pm$ are also constants (see eq. \eqref{Beta=dlog}).
Finally, $f(x)$ can be determined  by using (\ref{relatingvielbein}) in the second expression of (\ref{bosonicAVZ}). Comparison with (\ref{integrability}) leads to the following conditions for $f(x)$:
\begin{align}
df+\beta \,f&=0\,,\label{eqf0}
\\
f\,\tau&=\frac{1}{2}\,(\bar{\chi}_+\chi_++\bar{\chi}_-\chi_-)\,.\label{eqf}
\end{align}
Comparing eq. \eqref{eqf0} with \eqref{Beta=dlog}, and using \eqref{betatau}, it is easy to see that eq. \eqref{eqf0} is satisfied by $f=\alpha_\pm \,\tau_\pm$, where $\alpha_\pm$ are  {constants}. Note that the Bianchi identities for $R^i_\pm$, in eqs. (\ref{r1}), imply
\begin{equation}
\mathcal{D}[\Omega_\pm]R^i_\pm=0\;\;\,\Rightarrow\;\;\,
d(\bar{\chi}_\pm \chi_\pm)= -2\, \beta\,\bar{\chi}_\pm \chi_\pm\,.\label{dchichi}
\end{equation}
Then,  in a local patch where $\beta=0$, $\bar{\chi}_\pm \chi_\pm$ are separately  constant.

Let us now turn to the discussion of the fermionic sector of the model.
Restricting to the case $\beta =0$ (and then $\tau_\pm$ constant),
from the last two of eqs. (\ref{AT-likecurv}) one find, after some algebra,  the following Dirac equations:
\begin{equation}\label{Diracs}
 \slashed{\mathcal{D}}[\Omega_\pm,\,A_\pm]\,\chi_\pm=-3\,i\,\tau_\pm\,\chi_\pm\,,
\end{equation}
 where
$
\mathcal{D}[\Omega_+,\,A_+]\,\chi^{a_1}_+ \equiv d\chi^{a_1}+\frac{i}{2}\,\Omega^i_+\,\gamma_i\chi^{a_1}+
A^{a_1b_1}\chi_{b_1}
$,
and similarly for $\mathcal{D}[\Omega_-,\,A_-]\chi_-$.
Equations (\ref{bosonicAVZ}) and (\ref{Diracs}) can  be derived as Euler-Lagrange equations from the Lagrangian \eqref{Lagrangianpm}, by using the matter Ansatz \eqref{gravchiext}.

\subsubsection{{Relating torsion components to possible mass parameters}}\label{FNYW}
In the following, we will generally assume to have fixed the NYW symmetry  to set $\beta=0$, at least locally in an open neighborhood of the world volume.  Then, one can  use the remaining global NYW symmetry to fix either $\tau_+$ or $\tau_-$ (which are constants) to some chosen value.
It is useful to write the field equations of the spinors, eq. \eqref{Diracs}, in terms of the torsion-free Lorentz connection
$\omega^{\prime i}$
\begin{equation}
\omega^{\prime i}={\Omega_+^i}+\tau_+\,e^i={\Omega_-^i}+\tau_-\,e^i\,.
\end{equation}
Then, the Dirac equations in the two sectors can be recast in the form
\begin{equation}\label{Diracs2}
 \slashed{\mathcal{D}}[\omega^{\prime },\,A_\pm]\chi_\pm=-\frac{3}{2}\,i\,\tau_\pm\,\chi_\pm\,,
 \end{equation}
where, as in the models discussed in \cite{Alvarez:2011gd, Andrianopoli2018}, the mass of the spinor fields are fixed in terms of the torsion
\begin{equation}
    \label{diracmass}
m_\pm= \frac32\, \tau_\pm\,.
\end{equation}
The Riemann tensor associated with $\omega^{\prime }$, using eq.s (\ref{bosonicAVZ}), (\ref{betatau}), (\ref{eqf0}) and (\ref{eqf}),  reads
 \begin{equation}\label{R'}
R^i[\omega^{\prime }]=\frac{1}{2}\left(\frac{f^2}{\ell^2}+\tau^2+\frac{\eta_{AB}
 \bar{\chi}^A\chi^B}{\ell}\right)\,\epsilon^{ijk} e_j\wedge e_k\,.
 \end{equation}
Since $\beta=0$, as we have seen the coefficient of $\epsilon^{ijk} e_j\wedge e_k$ {in (\ref{R'}) is a constant that} defines an effective cosmological constant, which also receives a contribution from the fermion condensate (recall that then $f$, $\tau$ and $\bar\chi\chi$ are all constant and related by eq.\ (\ref{eqf})). 
The residual global NYW symmetry can be used to identify the AdS$_3$ radius of the world volume, let us call it  $\ell'$, with the one on the fiber, $\ell$. This still allows for several choices of world-volume spin connection, which can be labeled by a real parameter $\lambda$. Identifying the gauge connection $\Omega^i_{(\lambda)}\equiv \omega^i+ \frac{\lambda}{\ell}\,E^i$ with the tangent space connection $\Omega^{\prime i}_{(\lambda)}\equiv \omega^{\prime i}+ \frac{\lambda}{\ell}\,e^i$, yields
\begin{equation}
\tau=\frac{\lambda}{\ell}\,(f-1) \:,
\end{equation}
which combined with (\ref{betatau}) gives
\begin{equation}
    \tau_\pm=\frac{1}{\ell}\left[ \lambda\,(f-1)\mp f\right]\,.\label{taupmlambda}
\end{equation}
%
%
In this case eq.\ (\ref{eqf}) implies
\begin{equation}
\lambda\,f\,{(f-1)}=\frac{{\ell}}{2}\,(\bar{\chi}_+\chi_++\bar{\chi}_-\chi_-)\,.\label{ffm1p3}
\end{equation}
Under the parity  reflection discussed in Section \ref{parity}, $\lambda$ changes sign, so that parity invariance requires $\lambda=0$, which in turn implies $\bar{\chi}_+\chi_+=-\bar{\chi}_-\chi_-$.

In view of our  application, in the next Section, to the effective description of graphene-like systems, where parity will play an important role, we will only consider here  the parity-invariant  choice $\lambda=0$, corresponding to the A-T supergravities with $p=q$.  More general cases are discussed in \cite{Andrianopoli:2019sip}.
We will restrict in particular to the case $p=q=2$, where $a_1=a_2=1,2$.
In this special case, the 2 couples of Majorana spinors that we labeled as $\chi_+\equiv\chi_{a_1}$ and $\chi_-\equiv\chi_{a_2}$, can be collected in two Dirac spinors:
\begin{align}\label{dirac}
    \chi_{(+)}\equiv \chi_{a_1=1}+ i \chi_{a_1=2}\,,\quad \chi_{(-)}\equiv \chi_{a_2=1}+ i \chi_{a_2=2}\,.
\end{align}


\subsection{Few words about Graphene and 2D materials}\label{4}

A graphene sheet is a two-dimensional system of carbon atoms arranged in a honeycomb lattice \cite{Novoselov66}. \footnote{The literature on the subject is huge, and I refrain to cite all the important contributions here. For a comprehensive review, see \cite{neto2009electronic} and references therein. An extended list of references, focused on our analog approach, can also be found in \cite{Andrianopoli:2019sip}.}. It is a non-relativistic system which however provides a real framework to study Dirac pseudoparticles at sub-light speed regime, in an analog model where the Fermi velocity $v_F$ plays the role ot the speed of light $c$. 
%
The Dirac spinorial formulation emerges from the peculiar  honeycomb structure of graphene, where a \emph{unit cell} is made of two adjacent atoms belonging to topological  inequivalent sublattices, labelled A and B, respectively. The single-electron wave function is then conveniently described as a two-component Dirac spinor, $\zeta$ which 
can be written, in a proper basis of $\gamma$-matrices,
as \begin{equation}
\zeta=\left(\begin{array}{c}
    \sqrt{n_{A}}\, e^{i\alpha_A} \\
    \sqrt{n_{B}}\, e^{i\alpha_B}
\end{array}\right)\,, \label{zetaAB}
\end{equation}
where $n_{A}$, $n_{B}$ are the probability densities for the electrons in the $\pi$-orbitals, referred to the A and B sublattices respectively, and $\alpha_A,\alpha_B$ the corresponding wave-function phases. In terms of $\zeta$,  the following two physical quantities can be defined:
\begin{equation} \label{n-dn}
n\equiv n_A+n_B=\zeta^\dagger \zeta\,,\qquad
\Delta n\equiv n_B-n_A= \bar{\zeta}\zeta\equiv \zeta^\dagger\gamma^0 \zeta\,,
\end{equation}
where $n$ is the total electron probability density while $\Delta n$
is the asymmetry in the probability density of finding the electron in one or the other of the two sublattices. This description is robust under changes of the lattice preserving the topological structure.

The Dirac physics is realized for low-lying energy quasiparticle excitations: for energy ranges where the electron wavelength is much larger than the lattice length,  the graphene lattice looks to the charge carriers as a continuum 2  dimensional space. Under these conditions, the wave function \eqref{zetaAB} behaves as a quantum Dirac field  in curved space \cite{Boada:2010sh,Gallerati:2018dgm}.

The Dirac spinor $\zeta$ can  be related to our spinors $\chi_\pm$  as follows:
\begin{equation}
\chi=\sqrt{\frac{\ell}{2}}\,U\,\zeta\,, \label{chizeta}
\end{equation}
where now, recalling \eqref{dirac}, $\chi$ is the   Dirac spinor
 $  \chi\equiv \chi_{(+)}+ \chi_{(-)}$,
while $U$ is the $2\times 2$ unitary matrix $U=\frac{1}{\sqrt{2}}\left(\begin{matrix}1 & 1 \cr -i & i\end{matrix}\right)$.
An important consequence of the relation (\ref{chizeta}), as first shown in \cite{Andrianopoli2018}, is that the quantity \begin{equation}\label{chichinanb}
\bar{\chi} \chi =\frac{\ell}{2}\,\bar{\zeta} \zeta=\frac{\ell}{2}\,(n_B-n_A)\,
\end{equation}
  is constant for $\beta=0$ by virtue of the last of eqs. (\ref{dchichi}), in which case the difference $n_B-n_A$ is a constant index.
In our case (recalling eq.(\ref{ffm1p3})),  we have:
\begin{align}
    \lambda=0 \,: \bar\chi\chi=0 \quad \Rightarrow \quad n_A=n_B\,,
\end{align}
where
$n_A=n^{(+)}_A+n^{(-)}_A$; $
n_B=n^{(+)}_B+n^{(-)}_B$, and
$n^{(\pm)}_A$, $n^{(\pm)}_B$ denote the probability densities in the $+$ and $-$ sectors of the A and B sublattices.
Note that  the condition $\bar{\chi}_+\chi_+=-\bar{\chi}_-\chi_-$, and $n_A=n_B$,} can be realized as a non-trivial relation between the probability densities in the $+$ and $-$ sectors.



\subsubsection{The    {$\bf{K}$, $\bf{K'}$} Dirac points}

%
%
The electronic band structure of graphene, close to the Fermi energy, has a linear energy-momentum relationship, known as a Dirac cone, resembling the one of a relativistic spinor particle.
In this energy range, the Hamiltonian has two eigenvalues with opposite signs, corresponding to the conduction ($>0$) and valence ($<0$) bands of graphene. The two bands touch each other at the conical apices, known as \emph{Dirac points}, which define the \emph{reciprocal lattice}. The reciprocal lattice of graphene also features a honeycomb geometry where,  similarly to the primitive lattice,  the unit cell is made of two inequivalent points, $\mathbf{K}$ and  $\mathbf{K'}$, referred to as the 2 “valleys". As a result, electrons in graphene possess the \emph{valley} ($\mathbf{K}$ or $\mathbf{K'}$) as an additional pseudo-spin number, besides the site number $A$ or $B$ (a comprehensive discussion can be found in the review \cite{neto2009electronic}).

The $\pm$ sectors of A-T supergravity in the $p=q$ case, which are interchanged by a reflection symmetry in one spatial axis, can be naturally associated with the ${\bf K},\,{\bf K}'$ valleys of graphene.
To clarify how the parity symmetry emerges in the graphene analog, we still have to extend the action of parity to include the matter Ansatz \eqref{gravchiext}:
\\
First of all,  invariance of the full supersymmetric model requires eq. (\ref{parity2}), that in turn implies, using \eqref{gravchiext}:
\begin{align}
\chi_{(\pm)}\,\rightarrow\, \tilde{\chi}_{(\pm)}=-\sigma^1\chi_{(\mp)}\,.
\label{parity3}
\end{align}
Moreover,
given the relation $E^i=f\,e^i$ implied by  \eqref{gravchiext}, assuming the constant parameter $f$ to be parity-invariant, the action of parity  on $E^i$ naturally extends to $e^i$:
\begin{equation}
    e^i\,\rightarrow\, \tilde{e}^i=\mathcal{O}_Y{}^i{}_j\,e^j\,.\label{parity15}
\end{equation}
Invariance under $Y$-parity reflection of the expression (\ref{integrability}) for torsion requires: $\tau_+=-\tau_-$.
%
The field equations are parity invariant, while the Lagrangian density is odd. The two Dirac equations (\ref{Diracs2}) are mapped into one another (see \cite{Andrianopoli:2019sip} for details).

In the analog model, the same parity-reflection  exchanges the A and B sites. If the reflection is combined with a time-reversal transformation, the valleys ${\bf K}$ and ${\bf K}'$ are mapped into each other, so that, on a momentum vector: ${k_x\rightarrow k_x\,,\;k_y\rightarrow -k_y}$.
\\In the absence of curvature, the Dirac equations in momentum space in the two valleys, in our conventions, and setting $\hbar=v_\textsc{f}=1$, read \cite{neto2009electronic}:
\begin{align}
\begin{split}
&{\bf K}\: :\qquad E_{{\bf q}}
\;\chi_{{}_{\bf K}}({\bf q})=\left[\gamma^0\left(\gamma^1\,q^1+\gamma^2\,q^2+ m_{{}_{\bf K}} \right)\right]\chi_{{}_{\bf K}}({\bf q})\;,
\\
&{\bf K}':\qquad E_{{\bf q}}
\;\chi_{{}_{{\bf K}'}}({\bf q})=\left[\gamma^0\left(\gamma^1\,q^1-\gamma^2\,q^2+ m_{{}_{{\bf K}'}} \right)\right]\chi_{{}_{{\bf K}'}}({\bf q})\;,
\label{KKp}
\end{split}
\end{align}
where the two equations hold for the two-momenta ${\bf K}+{\bf q}$ and ${\bf K}'+{\bf q}$, with ${|{\bf q}|\ll |{\bf K}|,  |{\bf K}'|}$.
In configuration space, they read:
\begin{align}
\begin{split}
&{\bf K}\: :\qquad i\,\partial_t\chi_{{}_{\bf K}}({  x})=\,\left[\gamma^0\left(-i \gamma^1\,\partial_x-i \gamma^2\,\partial_y+ m_{{}_{\bf K}} \right)\right]\chi_{{}_{\bf K}}({x})\;,
\\
&{\bf K}':\qquad i\,\partial_t
\chi_{{}_{{\bf K}'}}(x)=\left[\gamma^0\left(-i \gamma^1\,\partial_x+i \gamma^2\,\partial_y+ m_{{}_{\bf K}'} \right)\right]\chi_{{}_{{\bf K}'}}(x)\;.
\label{KKpx}
\end{split}
\end{align}
By general covariance,  eq.s (\ref{KKpx}) can be extended to a curved background, with spinors minimally coupled to a gauge potential, by replacing partial derivatives with covariant ones ($\nabla^{\mathbf{K}}_\mu\equiv \mathcal{D}_\mu[\omega',\,A_{{}_{\bf K}}]$, $\nabla^{\mathbf{K}'}_\mu\equiv \mathcal{D}_\mu[\omega',\,A_{{}_{\bf K'}}]$):
\begin{align}
\begin{split}
{\bf K}:\quad &i \,\nabla^{\mathbf{K}}_t\,\chi_{{}_{\bf K}}(x)=\,\left[-i \gamma^0\left(\gamma^1\,\nabla^{\mathbf{K}}_x+\gamma^2\,\nabla^{\mathbf{K}}_y+ i m_{{}_{\bf K}}\right)\right]\chi_{{}_{\bf K}}(x)\;,
\\
{\bf K'}:\quad &i \,\nabla^{\mathbf{K}'}_t\,\chi_{{}_{\bf K}'}(x)=\,\left[-i \gamma^0\left(\gamma^1\,\nabla^{\mathbf{K}'}_x-\gamma^2\,\nabla^{\mathbf{K}'}_y+ i m_{{}_{\bf K}'}\right)\right]\chi_{{}_{\bf K}'}(x)\;,
\label{KKpxD}
\end{split}
\end{align}
where $A_{{}_{{\bf K}}}(x)$ and $A_{{}_{{\bf K}'}}(x)$ denote the gauge fields about the two Dirac points.

By comparing (\ref{KKpxD}) with eqs. (\ref{Diracs2}),
we can consistently identify the spinor fields $\chi_\pm(x)$ with $\chi_{{}_{\bf K}}(x),\,\chi_{{}_{{\bf K}'}}(x)$, up to an overall normalization, as follows
\begin{equation}
\chi_{{}_{\bf K}}(x^\mu)= \chi_+(x^\mu)\;,\qquad
\chi_{{}_{{\bf K}'}}(x^0,\,x^1,\,x^2)= \sigma^1\chi_-(-x^0,\,x^1,\,x^2)\;,
\end{equation}
provided we also identify the gauge fields as the $\mathrm{SO}(2)_\pm$ ones of the A-T model:
\begin{equation}
A_{{}_{{\bf K}}}=A_+\,,\quad\;\;
A_{{}_{{\bf K}'}}=A_-\,,
\end{equation}
and the mass gaps at the two valleys with the mass parameters $m_\pm$ of $\chi_\pm$:
\begin{equation}
m_{{}_{\bf K}}=m_+=\frac{3}{2}\,\tau_+\;,
\qquad
m_{{}_{{\bf K}'}}=m_-=\frac{3}{2}\,\tau_-\;.
\label{midentification}
\end{equation}
This motivates the identification of the $\pm$ sectors in our model with the two valleys, and the corresponding mass gaps with the torsional parameters of our model.
Note that applying spatial reflection with respect to the X-axis maps eqs.\ (\ref{KKp}) into each other, provided $m_{{}_{{\bf K}'}}=-m_{{}_{\bf K}}$\,. This implies that $m_{{}_{{\bf K}'}}+m_{{}_{\bf K}}$ is parity-odd, while $m_{{}_{{\bf K}'}}-m_{{}_{\bf K}}$ is parity-even.
\subsubsection{{Microscopic interpretation}}
Mass terms can be included in graphene by generalizing the tight binding microscopic model  and opening mass gaps at the Dirac points.
I will skip here all the details, some of which can be found in \cite{Andrianopoli:2019sip} (see in particular Section 4 and Appendix B of that paper).
The generation of a gap at Dirac points was first discussed in 1984 by Semenoff, introducing a mass term through an on-site  deformation, $\pm M$, breaking the sublattices equivalence \cite{Semenoff:1984dq}.
Another model, which includes a periodic local magnetic flux density with zero net flux over the honeycomb hexagon  was proposed by Haldane \cite{Haldane:1988zza}.  The local fluxes on the cells induce  an Aharonov–Bohm  phase, $\varphi$,  which can be taken as a parameter of the model.
Either a non-zero Semenoff mass $M$ or Aharanov-Bohm phase $\varphi$ lift the degeneracy of the bands at the Dirac points, and the fermion masses in the two inequivalent valleys turn out to be
\begin{equation}
m_{{}_{\bf K}}=M- 3\sqrt 3\, t_2 \sin{\varphi}\;,\qquad
m_{{}_{{\bf K}'}}=M+ 3\sqrt 3\, t_2 \sin{\varphi}\;,
\label{sh}
\end{equation}
where the coefficient $t_2$ is a  parameter of the  Haldane model \cite{Haldane:1988zza}.

Using eq.\ (\ref{midentification}) the physical quantities expressed by the Semenoff local potential term $M$ and Haldane contribution $3\sqrt{3}\,t_2\sin{\varphi}$ {can be related} to the fermion masses $m_\pm$ of our macroscopic model. To see this, let us recall eqs.\ (\ref{betatau}), that implies:
\begin{equation}
\tau_\pm \,=\,\tau \mp\,2\,\frac {f}{\ell}\,.\label{mpmtpm}
\end{equation}
 The  first term is parity odd 
 while the second is parity even. Recalling eq.s (\ref{midentification}), this suggests  the following identification:
\begin{equation}
M=\frac 32\tau\;,\quad\;\;
\sqrt 3\,t_2 \sin(\varphi)=\frac{ f}{\ell}\;.
\label{identification}
\end{equation}

\section{Conclusions and outlook}
The spinorial model obtained after this long journey, which started from $\mathcal{N}=2$ supergravity in $D=4$ to end up with a $2+1$-dimensional sheet of graphene, has the nice feature of embedding, in a top-down approach, fields and parameters of graphene-like materials in a $D=4$ supergravity model, where the wave function for the electron probability in the bipartite unit cell of graphene is interpreted as the asymptotic value of the radial component of the gravitino.
It is an analog relationship, whose details are still largely to be explored. Among many others, there is a specific feature that deserves deepening and that I would like to mention:
The complex wave function of the electrons in graphene-like materials is \commas relativistic" because it is bipartite, as a 2-component Dirac spinor does in $D=3$, but also because,  close to the Dirac points, this non-relativistic wave function satisfies eq. \eqref{KKpxD}, namely a (analog) Dirac equation. In that relation,   the Fermi velocity $v_{F}$ (hidden in the choice of  \commas Natural units") plays the role of analog speed of light. However, in our top-down approach the speed of light comes from the $D=4$ supergravity model, and is naturally identified with the actual speed of light, $c$.
A possible way to deal with this issue is to define a more general, matricial relation between the two sets of dreibein $E^i$ and $e^i$,  than the one assumed here in (\ref{relatingvielbein}), that is of the form $ E^i={M}^i_{\,\,j}\,e^j$. By choosing, for instance, ${M}^i_{\,\,j}\propto {\mathrm{diag}}(\alpha^2,1,1)$, one can introduce  in the world-volume an analog speed of light $\hat c=\alpha \,c$.
The mathematical implications of such assumption were investigated in \cite{Noris:2019sdw}. The inclusion of the torsion component $\mathring{T}^{ij}$ in the decomposition of the $3D$-torsion tensor, eq. \eqref{rescal}, could also be useful, since it would combine with the above generalization to allow setting two independent scales of velocities. This issue, together with the construction of explicit solutions taking into account the matter Ansatz \eqref{gravchi} in backgrounds including topological non-trivial configurations and the effects of  Beltrami-torsion, is currently  under investigation.

Another aspect to be explored is the formulation of the above described relationship in terms of holographic renormalization, to clarify if it is possible to express it as a proper gauge/gravity duality. A preliminary result is the holographic renormalization analysis of \cite{Andrianopoli:2020zbl}, which includes all the fermionic contributions, and is ready to be applied to to the unconventional scenario described here.
\section*{Acknowledgements}
This review collects the main results of several papers, which were obtained in various collaborations with many friends, 
in particular   B.L. Cerchiai, R. D'Auria, A. Gallerati, P.A. Grassi, R. Matrecano, O. Miskovic, R. Noris, R. Olea,
L. Ravera, M. Trigiante and J. Zanelli.  I am grateful to all of them.

\end{document}